\documentclass[10pt, final, journal, letterpaper, twocolumn]{IEEEtran}
\makeatletter
\def\ps@headings{%
	\def\@oddhead{\mbox{}\scriptsize\rightmark \hfil \thepage}%
	\def\@evenhead{\scriptsize\thepage \hfil \leftmark\mbox{}}%
	\def\@oddfoot{}%
	\def\@evenfoot{}}
\makeatother 

\IEEEoverridecommandlockouts
\usepackage{bbm}
\usepackage{amsfonts}
\usepackage[dvips]{graphicx}
\usepackage{times}
\usepackage{cite}
\usepackage{amsmath}
\usepackage{array}
\usepackage{amssymb}

\usepackage{stfloats}
\usepackage{graphicx}
\usepackage{footnote}
\usepackage{booktabs}
\usepackage{array}
\usepackage{algorithm}
\usepackage{subeqnarray}
\usepackage{cases}
\usepackage{threeparttable}
\usepackage{color}
\usepackage{hyperref}
\usepackage{epstopdf}
\usepackage{algpseudocode}
\usepackage{bm}
\usepackage{multirow}
\usepackage[labelformat=simple]{subcaption}
\usepackage{adjustbox}
\usepackage{cuted}

\newtheorem{theorem}{Theorem}

\newtheorem{lemma}{Lemma}

\newtheorem{corollary}{Corollary}

\begin{document}

\title{Ergodic Achievable Rate Analysis and Optimization of RIS-assisted Millimeter-Wave MIMO Communication Systems}

\author{\authorblockN{Renwang Li, Shu Sun, \IEEEmembership{Member,~IEEE}, Yuhang Chen, Chong Han, \IEEEmembership{Member,~IEEE}, Meixia Tao, \IEEEmembership{Fellow,~IEEE}}\\
\thanks{This work is supported in part by the National Key R\&D Project of China under grant 2020YFB1406802 and the National 	Natural Science Foundation of China under grant 61941106. (Corresponding author: Meixia Tao.)}
\thanks{R. Li, S. Sun, and M. Tao are with Department of Electronic Engineering, Shanghai Jiao Tong University, Shanghai, China (emails:\{renwanglee, shusun, mxtao\}@sjtu.edu.cn).}
\thanks{Y. Chen is with Terahertz Wireless Communications (TWC) Laboratory, Shanghai Jiao Tong University, Shanghai, China (emails:\{yuhang.chen\}@sjtu.edu.cn).}
\thanks{C. Han is with Terahertz Wireless Communications (TWC) Laboratory and Department of Electronic Engineering, Shanghai Jiao Tong University, Shanghai, China (emails:\{chong.han\}@sjtu.edu.cn).}
}

\maketitle

\begin{abstract}
Reconfigurable intelligent surfaces (RISs) have emerged as a prospective technology for next-generation wireless networks due to their potential in coverage and capacity enhancement. Previous works on achievable rate analysis of RIS-assisted communication systems have  mainly focused on the rich-scattering environment where Rayleigh and Rician channel models can be applied. This work studies the ergodic achievable rate of RIS-assisted multiple-input multiple-output communication systems in millimeter-wave band with limited scattering  under  the Saleh-Valenzuela channel model. Firstly, we derive an upper bound of the ergodic achievable rate by means of majorization theory and Jensen's inequality.  The upper bound shows that the ergodic achievable rate increases logarithmically with the number of antennas at the base station (BS) and user, the number of the reflection units at the RIS, and the eigenvalues of the steering matrices associated with the BS, user and RIS. Then, we aim to maximize the ergodic achievable rate by jointly optimizing the transmit covariance matrix at the BS and the reflection coefficients at the RIS. Specifically, the transmit covariance matrix is optimized by the water-filling algorithm and the reflection coefficients are optimized using the Riemannian conjugate gradient  algorithm. Simulation results validate the effectiveness of the proposed optimization algorithms.
\end{abstract}

\begin{IEEEkeywords}
Reconfigurable intelligent surface, achievable rate, statistical channel state information (CSI), transmit covariance matrix, reflection coefficients, millimeter-wave communications.
\end{IEEEkeywords}

\section{Introduction}
The millimeter-wave (mmWave) communication over the 30-300 GHz spectrum is regarded as a promising technology for 5G and beyond wireless networks because of its great potential to offer high communication data rates \cite{6824746, niu2015survey}. Large antenna arrays are usually adopted in mmWave communications since they can form highly directional beams to compensate for the severe path loss compared with sub-6 GHz. Meanwhile, the high directivity makes mmWave communication much more sensitive to signal blockage \cite{8254900}.

To overcome the blockage issue, reconfigurable intelligent surfaces (RISs) have been recently introduced to enlarge the coverage of the mmWave communication systems \cite{8910627,9148781}. Specifically, an RIS is an artificial meta-surface that is composed of a large number of  passive units and can be  controlled efficiently by a smart controller \cite{8796365}. An RIS can smartly reflect the signals from a base station (BS) to multiple users by adjusting the amplitude and phase of the incident signals. When the channel between the BS and the user is blocked, an RIS can create a concatenated BS-RIS-user channel, and thus improve the coverage of mmWave systems. There are often no radio frequency (RF) components on RIS. Therefore, RIS is an efficient and cost-effective solution for the blockage problem in mmWave systems.

Motivated by the above attractive advantages, RIS-aided communication systems have been widely studied in various scenarios, e.g., \cite{8811733,8982186,8723525, 9110912, 9234098}. The authors in \cite{8811733} consider the power minimization problem for an RIS-assisted multiple-input single-output (MISO) downlink multi-user system subject to signal-to-interference-plus-noise ratio  constraints. The weighted sum-rate is maximized in \cite{8982186} under the maximum power budget. The authors in \cite{8723525} focus on maximizing the secrecy rate of the legitimate user. The authors in \cite{9110912} aim to maximize the capacity of an RIS-aided multiple-input multiple-output (MIMO) system and propose an alternating optimization based algorithm. Pertaining to mmWave MIMO systems, the spectral efficiency maximization problem is investigated in \cite{9234098} by exploiting the sparse structure of mmWave channel to propose an efficient manifold-based algorithm.

However, all of the above contributions are based on instantaneous channel state information (CSI), which is very challenging to obtain due to the fact that RIS is usually a passive component and cannot sense the environment independently \cite{8910627, 8796365}.  Thus, some researchers attempt to explore the statistical CSI in RIS-aided communication systems \cite{9195523, singh2021reconfigurable, 8746155,hu2020statistical,9371709, zhi2021power,9500188, 9352967,  9392378  }. For single-input single-output  systems,  \cite{9195523} considers the Rayleigh fading channel and obtains a closed-form expression of coverage probability. The Rician channel is considered in \cite{singh2021reconfigurable}. For MISO systems,  \cite{8746155} derives a tight upper bound of the ergodic spectral efficiency under Rician fading channels. Therein, the maximum ratio transmission  is adopted at the BS and  the optimal reflection coefficients at the RIS are designed based on the upper bound.  In \cite{hu2020statistical}, the authors take the active beamforming at the BS into consideration when deriving the achievable rate. {The authors in \cite{9371709} detect the angles at the BS by maximum likelihood estimators, and then propose  joint optimization of BS beamforming and RIS beamforming based on the estimated angles. The work in  \cite{8746155},  \cite{hu2020statistical}, and \cite{9371709} all consider the single user scenario. The extension to the multiuser case is considered in  \cite{zhi2021power,9500188}.} For MIMO systems,  \cite{9352967} considers single-stream transmission and obtains a tight upper bound of the ergodic spectral efficiency under the Rician channels, while the multi-stream transmission case is considered in  \cite{9392378}.  Note that all these works consider the Rayleigh or Rician channels. However, the Rayleigh or Rician channels are not suitable for mmWave systems due to the limited scattering. Instead, mmWave communication channels are usually characterized by the Saleh-Valenzuela (SV) channel \cite{9234098,6847111,6717211,7397861,7962632,8816689}. {To our best knowledge, there is no existing work on the analysis of the ergodic achievable rate under the SV channel in RIS-aided mmWave MIMO communication systems.}

Against the above background, we concentrate on  the achievable rate analysis and optimization of an RIS-aided downlink mmWave MIMO communication systems with the  statistical CSI under SV channel model. The main contributions of this study are summarized as follows:
\begin{itemize}
  \item We derive a closed-form upper bound of the ergodic achievable rate of the RIS-aided mmWave MIMO communication systems based on the majorization theory and Jensen's inequality. The upper bound is applicable to arbitrary numbers of paths, antennas, and reflection units, and arbitrary antenna structures.
      The upper bound shows that the ergodic achievable rate increases strictly with the number of antennas at the BS and user, the number of the reflection units at the RIS, and the eigenvalues of the steering matrices associated with the BS, user and RIS.
  \item We jointly optimize the transmit covariance matrix at the BS and the reflection coefficients at the RIS based on the derived closed-form expression for ergodic achievable rate maximization. Specifically, the transmit covariance matrix is optimized by the water-filling algorithm, and the reflection coefficients are optimized by the Riemanian conjugate gradient (RCG) algorithm.
  \item We conduct comprehensive simulations to validate the performance of  the proposed algorithms. It is shown that the ergodic achievable rate after optimization can be improved by about 20 bits/s/Hz. In addition, when the transmit signals on each antenna are  independent and identically distributed (i.i.d.), the ergodic achievable rate remains unchanged after the number of antennas at the BS reaches a certain amount. Furthermore, it is found that when the RIS has a sufficient number of reflection units, optimizing the reflection coefficients at the RIS is more effective than optimizing the transmit covariance matrix at the BS. Meanwhile, the system still performs well if the BS allocates equal power over the transmit antennas and the reflection coefficients at the RIS are optimized.
\end{itemize}

{The rest of the paper is organized as follows. {Section II introduces the system model and the problem formulation. Section III adopts the majorization theory to derive the approximation for the ergodic achievable rate. Section IV optimizes the ergodic achievable rate by designing the  transmit covariance matrix and the reflection coefficients.} Section V presents the extensive simulation results and Section VI concludes this paper.}

\emph{Notations}: The imaginary unit is denoted by $j=\sqrt{-1}$. Vectors and matrices are denoted by bold-face lower-case and upper-case letters, respectively. $\mathbb{C}^{x\times y}$ denotes the space of $x\times y$ complex-valued matrices. $ \bf x^*$, $\mathbf{x}^T$, and $\mathbf x^H$ denote the conjugate, transpose and  conjugate transpose of vector $\bf x$. $\bf I$ denotes an identity matrix of appropriate dimensions. $\odot$ denotes the Hadamard product. $\mathbb{E}(\cdot)$ denotes the stochastic expectation. $\Re(\cdot)$ denote the real part of a complex number. The $\operatorname{tr}(\cdot)$, $\operatorname{det}(\cdot)$ and $\operatorname{rank}(\cdot)$ denote the trace, determinant and rank operation, respectively. $\operatorname{diag}(\mathbf{x})$ denotes a diagonal matrix with each diagonal element being the corresponding element in $\mathbf{x}$. $\nabla f(\mathbf{x}_i)$ denotes the gradient vector of function $f(\mathbf{x})$ at the point $\mathbf{x}_i$. The distribution of a circularly symmetric complex Gaussian  random vector with mean vector $x$ and covariance matrix $\Sigma$ is denoted by $\mathcal{C}\mathcal{N}(x,\Sigma)$; and $\sim$ stands for ``distributed as''. The exponential random variable ${X}$ with parameter $\lambda$ is given by ${X} \sim \exp (\lambda)$.

\section{System Model and Problem Formulation}
\label{sec_2}
\subsection{System Model} \label{sec_system}
\begin{figure}[t]
\begin{centering}
\includegraphics[width=.5\textwidth]{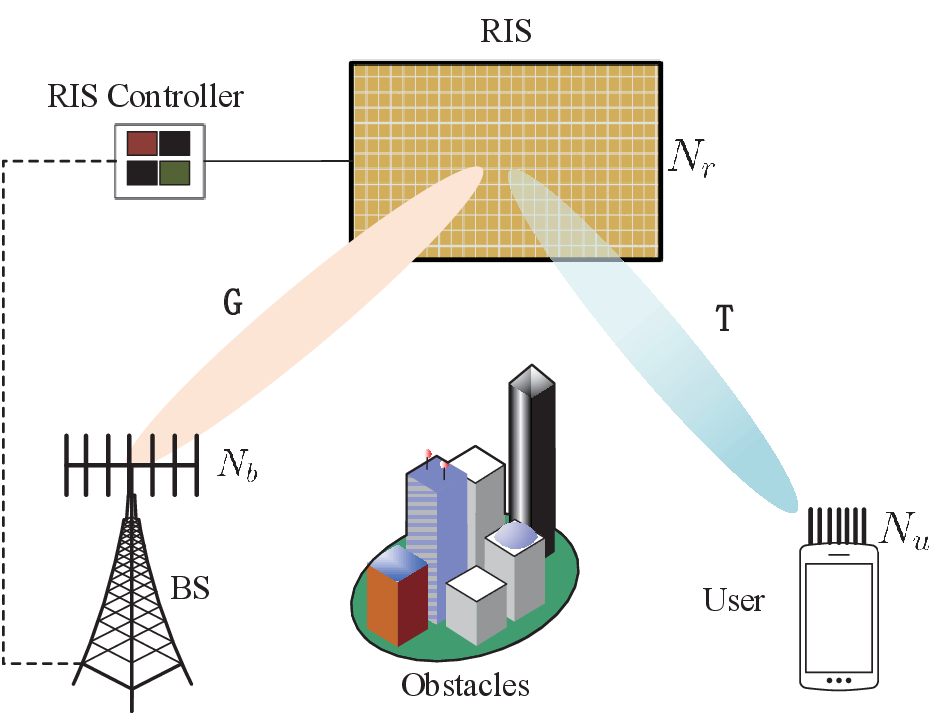}
 \caption{Illustration of an RIS-aided mmWave MIMO system.} \label{system_model}
\end{centering}
\vspace{-0.4cm}
\end{figure}

As shown in Fig.~\ref{system_model}, we consider an RIS-aided downlink mmWave MIMO communication system, where one BS equipped with $N_b\geq 1$ antennas communicates with a user equipped with $N_u\geq 1$ antennas, via the assistance of one RIS equipped with $N_r\geq 1$  passive reflection units. We assume that the BS-user direct link is blocked due to unfavorable propagation conditions. 
Let $\mathbf{G}\in\mathbb{C}^{N_r \times N_b}$ and $\mathbf{T}\in\mathbb{C}^{N_u \times N_r}$ denote the channel matrices from BS to RIS, and from RIS to user, respectively, and let $\mathbf{s} \in \mathbb{C}^{N_b \times 1}$ denote the zero-mean transmitted Gaussian signal vector with covariance matrix $\mathbf{Q}\in \mathbb{C}^{N_b \times N_b}$. The transmit power constraint at the BS can be expressed as
 \begin{equation}\label{budgetP}
\operatorname{tr} \mathbb{E}\{\mathbf{s}\mathbf{s}^H\} = \operatorname{tr}(\mathbf{Q} ) \leq P_T ,
\end{equation}
where $P_T>0$ is the power budget at the BS. Then, the received vector $\mathbf{y}\in \mathbb{C}^{N_u \times 1}$ at the user can be represented as
\begin{equation}
\mathbf{y}=\mathbf{T}\mathbf{\Theta}\mathbf{G} \mathbf{s} + \mathbf{n},
\end{equation}
where $\mathbf{\Theta}= \operatorname{diag} \{\xi_1 e^{j\theta_1}, \xi_2 e^{j\theta_2}, \ldots, \xi_{N_r} e^{j\theta_{N_r}}\}$ represents the response matrix of the RIS with $\theta_i\in [0, 2\pi)$ and $\xi_i \in[0, 1]$ being the phase shift and the amplitude reflection coefficient of the $i$-th reflection unit, respectively, and $\mathbf{n} \thicksim\mathcal{C}\mathcal{N}{(\mathbf{0}, \sigma^2 \mathbf{I}_{N_u})}$ represents the additive white Gaussian noise with zero mean and variance $\sigma^2$. In this paper, we assume $\xi_i=1, i= 1,2,\ldots, N_r$ to maximize the signal reflection.

We adopt the widely used narrowband SV channel model for mmWave communications. Suppose uniform linear arrays (ULAs)\footnote{We assume the deployment of ULA for ease of exposition. The results can be extended straightforwardly to other antenna topologies, such as UPA and so on. } are equipped at the BS and the user, and a uniform planar array (UPA) is equipped at the RIS.  Hence, the narrowband BS-RIS channel $\mathbf{G}$ and  RIS-user channel $\mathbf{T}$ can be expressed as
\begin{equation}\label{channel_G}
\mathbf{G}= \sqrt{\frac{N_r N_{b}}{P}} \sum_{i=1}^{P} g_i \mathbf{a}_r(\phi_{r,i}^a, \phi_{r,i}^e) \mathbf{a}_b^H (\phi_{b,i}),
\end{equation}
\begin{equation}\label{channel_T}
\mathbf{T}= \sqrt{\frac{N_r N_{u}}{L}} \sum_{i=1}^{L} t_i \mathbf{a}_u(\psi_{u,i}) \mathbf{a}_r^H (\psi_{r,i}^a, \psi_{r,i}^e),
\end{equation}
where $P (L)$ is the number of paths between the BS and the RIS (the RIS and the user), $g_i \sim \mathcal{C}\mathcal{N} (0,1)$ ($t_i \sim \mathcal{C}\mathcal{N} (0,1)$)\footnote{In mmWave MIMO communication systems, it is reasonable to assume that the complex gains of different paths experience the same variances \cite{6847111,6717211,7397861}. Furthermore, the results are extendible to the case of different variances. } denotes complex channel gain of the $i$-th path, $\phi_{r,i}^a$ ($\psi_{r,i}^a$) and $\phi_{r,i}^e$ ($\psi_{r,i}^e$) denote the azimuth and elevation angles of arrival (departure) associated with the RIS, $\phi_{b,i}$ denotes the angle of departure (AoD) associated with the BS, $\psi_{u,i}$ denotes the angle of arrival (AoA) associated with the user, and $\mathbf{a}_i, i\in \{ b, r, u\}$ denote the normalized array response vectors. When an $N$-antenna ULA is employed, the array response vector is given by
\begin{equation} \label{ula}
\mathbf{a}(\phi) = \frac{1}{\sqrt{N}}\left[ 1, e^{j\frac{2\pi d}{\lambda} \sin(\phi)},  \cdots,  e^{j \frac{2\pi d}{\lambda} (N-1) \sin(\phi)} \right]^T,
\end{equation}
where $\lambda$ is the signal wavelength, $d$ is the antenna spacing which is assumed to be half wavelength, and $\phi$ denotes the AoA or AoD. For a UPA with $M=M_y \times M_z$ elements, the array response vector is given by
 \begin{equation} \label{upa}
 	\begin{aligned}
	\mathbf{a} \left(\phi^a, \phi^e\right)=&
	 \frac{1}{\sqrt{M}} \left[1, \ldots, e^{j \frac{2 \pi}{\lambda} d \left(m_y \sin \phi^a \sin \phi^e + m_z \cos \phi^e\right)}\right. , \\
	 &
	\left.\ldots, e^{j \frac{2 \pi}{\lambda} d \left((M_y-1) \sin \phi^a \sin \phi^e)+(M_z-1) \cos \phi^e\right)}\right]^{T},
	\end{aligned}
\end{equation}
where $d$ is the unit cell spacing which is assumed to be half a wavelength, while $\phi^a$ and $\phi^e$ denote the azimuth and elevation angles, respectively. Defining
$ \mathbf{A}_{rp} = \left[ \mathbf{a}_r(\phi_{r,1}^a, \phi_{r,1}^e), \ldots, \mathbf{a}_r(\phi_{r,P}^a, \phi_{r,P}^e) \right] \in\mathbb{C}^{N_r \times P}$,
$\mathbf{A}_{b} = \left[ \mathbf{a}_b(\phi_{b,1}), \mathbf{a}_b(\phi_{b,2}), \ldots, \mathbf{a}_b(\phi_{b,P}) \right] \in\mathbb{C}^{N_b \times P}$,
$\mathbf{G}_p = \sqrt{\frac{N_r N_{b}}{P}} \text{diag}( g_1, g_2, \ldots, g_P)$, the channel matrix $\mathbf{G}$ \eqref{channel_G} can be rewritten as
\begin{equation}\label{channelG}
\mathbf{G}= \mathbf{A}_{rp} \mathbf{G}_p \mathbf{A}_b^H.
\end{equation}
Defining $\mathbf{A}_{rl} = \left[ \mathbf{a}_r(\psi_{r,1}^a, \psi_{r,1}^e),  \ldots, \mathbf{a}_r(\psi_{r,L}^a, \psi_{r,L}^e) \right] \in\mathbb{C}^{N_r \times L}$,
$ \mathbf{A}_{u} = \left[ \mathbf{a}_u(\psi_{b,1}), \ldots, \mathbf{a}_u(\psi_{b,L}) \right] \in\mathbb{C}^{N_u \times L}$,
$ \mathbf{T}_l = \sqrt{\frac{N_r N_{u}}{L}} \text{diag}( t_1, t_2, \ldots, t_L)$,
the channel matrix $\mathbf{T}$ \eqref{channel_T} can be rewritten as
\begin{equation}\label{channelT}
\mathbf{T}=  \mathbf{A}_{u} \mathbf{T}_l \mathbf{A}_{rl}^H.
\end{equation}

For a given transmit covariance matrix $\mathbf{Q}$ and a given RIS response matrix $\mathbf{\Theta}$, the ergodic achievable rate of the RIS-aided mmWave MIMO communication systems can be expressed as
\begin{equation} \label{capacity_ori}
R(\mathbf{Q}, \mathbf{\Theta}) = \mathbb{E}_{\mathbf{H}}\left[\log _{2} \operatorname{det}\left(\mathbf{I}_{N_{u}}+\frac{1}{\sigma^{2}} \mathbf{H} \mathbf{Q H}^{H}\right)\right],
\end{equation}
where $\mathbf{H}=\mathbf{T}\mathbf{\Theta}\mathbf{G}$ denotes the cascaded BS-RIS-user channel. {The expectation is taken over the cascaded channel $\mathbf{H}$. More specifically, it is taken over the complex channel gain $\{g_i\}_{i=1,2,\ldots,P}$ in Eq. \eqref{channel_G} and $\{t_i\}_{i=1,2,\ldots,L}$ in Eq. \eqref{channel_T} in the SV channel model. }

\subsection{Problem Formulation}
We aim to maximize the ergodic achievable rate by jointly optimizing the transmit covariance matrix at the BS and the response matrix at the RIS, subject to the maximum power budget \eqref{budgetP} at the BS and the constant amplitude reflection constraints at the RIS. Therefore, the problem can be formulated as
    \begin{subequations}\label{prob_original}
    \begin{align}
            \mathcal{P}_0: \max \limits_{ \mathbf{Q}, \mathbf{\Theta} } \quad & R(\mathbf{Q}, \mathbf{\Theta}) \\
            \text { s.t. } \quad & \operatorname{tr}(\mathbf{Q}) \leq P_T, \label{const_q1}\\
            & \mathbf{Q} \succeq 0, \label{const_q2} \\
            &  \mathbf{\Theta} =\operatorname{diag}\left(e^{j \theta_{1}}, e^{j \theta_{2}}, \cdots, e^{j \theta_{N_r}} \right). \label{const_ris}
    \end{align}
    \end{subequations}
The problem $\mathcal{P}_0$ is challenging mainly due to two facts. To begin with, there is no explicit expression of the ergodic achievable rate, which prevents further optimization exploration. Second, the problem is highly non-convex due to the unit-modulus phase shifts constraints \eqref{const_ris}. In the following, we will first derive a closed-form upper bound of the ergodic achievable rate in Section \ref{sec_analysis}, and then find the optimal $\mathbf{Q}$ and $\mathbf{\Theta}$ to maximize the ergodic achievable rate in Section \ref{sec_optimization}.

\section{Analysis of the Ergodic Achievable Rate }\label{sec_analysis}
In this section, we first derive an approximation of the ergodic achievable rate according to the majorization theory. Then, an explicit expression is obtained via the Jensen's inequality. Finally, we evaluate the tightness of the approximations via numerical results.
\subsection{Approximation of the Ergodic Achievable Rate}
\begin{lemma} \label{lemma1}
Under the narrowband SV channel model expressed in Eq. \eqref{channelG} and Eq. \eqref{channelT}, the ergodic achievable rate $R(\mathbf{Q}, \mathbf{\Theta})$ in Eq. \eqref{capacity_ori} of the RIS-aided mmWave MIMO communication systems can be approximated by
\begin{equation} \label{app_ori}
	\begin{aligned}
R(\mathbf{Q}, \mathbf{\Theta})&\approx \widetilde{R}(\mathbf{Q}, \mathbf{\Theta}) \\
 &\triangleq \mathbb{E}_{g,t} \left[\sum \limits_{i=1}^{N_s} \log _{2} \left( 1+ \frac{ N_b N_u N_r^2 }{\sigma^2 P L } d_{b,i} d_{u,i} d_{r,i} |g_i|^2 |t_i|^2  \right) \right],
 \end{aligned}
\end{equation}
where $N_s= \min\left(\operatorname{rank}(\mathbf{A}_b^H \mathbf{Q} \mathbf{A}_b), \operatorname{rank}(\mathbf{A}_u^H \mathbf{A}_u), \operatorname{rank}(\mathbf{X}^H \mathbf{X})\right)$, $ \mathbf{X} = \mathbf{A}_{rl}^H \mathbf{\Theta} \mathbf{A}_{rp}$, and $( d_{b,1}, \ldots, d_{b, N_s})$, $(d_{u,1}, d_{u,2}, \ldots, d_{u,N_s})$ and $(d_{r,1}, d_{r,2}, \ldots, d_{r,N_s})$ are descending ordered eigenvalues of $\mathbf{A}_b^H \mathbf{Q} \mathbf{A}_b$, $\mathbf{A}_u^H \mathbf{A}_u$ and $\mathbf{X}^H \mathbf{X}$, respectively.
\end{lemma}
\begin{proof}
See Appendix \ref{append1}.
\end{proof}

{Notice that the approximate expression of the ergodic achievable rate depends explicitly on the channel gains $\{g_i\}$ and $\{t_i\}$ of the first $N_s$ paths in both the BS-RIS link and the RIS-user link and is irrelevant to the channel gains in other paths. Here, $N_s$ can also be viewed as the approximate rank of the cascaded BS-RIS-user channel.}

\begin{corollary} \label{app_exact}
When $\mathbf{A}_b$, $\mathbf{A}_u$, $\mathbf{A}_{rp}$ and $\mathbf{A}_{rl}$ are composed of the columns of unitary matrices, e.g., the discrete Fourier matrices, and the AoA at the RIS is symmetric to the AoD at the RIS, i.e., $\phi_{r,i}^a = \psi_{r,i}^a, \phi_{r,i}^e = \psi_{r,i}^e, \forall i$, $\mathbf{Q} = \frac{P_T}{N_b} \mathbf{I}_{N_b}$, and $\mathbf{\Theta}=\mathbf{I}_{N_r}$, the  approximation $\widetilde{R}(\mathbf{Q}, \mathbf{\Theta})$ is identical to the exact ergodic achievable rate, i.e.,
\begin{equation}
\widetilde{R}(\mathbf{Q}, \mathbf{\Theta}) = {R}(\mathbf{Q}, \mathbf{\Theta}).
\end{equation}
\end{corollary}

\begin{proof}
See Appendix \ref{proof_pro1}.
\end{proof}
\textbf{Corollary} \ref{app_exact} implies that the derived approximation exhibits tight performance when the transmit signals on each antenna are i.i.d., the RIS acts like a mirror, and all steering response matrices are taken from columns of unitary matrices. {Specifically, when the number of the antennas or the reflection units is very large, the steering response matrices $\mathbf{A}_b$, $\mathbf{A}_u$, $\mathbf{A}_{rp}$ and $\mathbf{A}_{rl}$ become asymptotically orthonormal matrices since the columns have the form as Eq. \eqref{ula} or Eq. \eqref{upa} \cite{6831723}.}

Given that all $\{g_i\}$ and $\{t_i\}$ are independent and follow the distribution of $\mathcal{C}\mathcal{N} (0,1)$, the  expectation in Eq. \eqref{app_ori} can be evaluated, which leads to the following Corollary.
\begin{corollary} \label{theorem2}
The approximation of the ergodic achievable rate in \textbf{Lemma} \ref{lemma1} can be expressed as
\begin{equation} \label{theorem2_eq}
\widetilde{R}(\mathbf{Q}, \mathbf{\Theta})
 = \frac{1}{ \ln 2} \sum \limits_{i=1}^{N_s} \int_{0}^\infty
e^{-z} e^{ \frac{1}{\alpha_i z} } E_1 \left(\frac{1}{\alpha_i z} \right) dz,
\end{equation}
where $\alpha_i=\frac{ N_b N_u N_r^2 }{\sigma^2 P L } d_{b,i} d_{u,i} d_{r,i}$, and $E_1(z)$ is the exponential integral function, defined as $E_1(z) = \int_1^\infty x^{-1} e^{-zx} dx, z>0$.
\end{corollary}

\begin{proof}
See Appendix \ref{appendix_theorem2}.
\end{proof}

\textbf{Corollary} \ref{theorem2} provides an exact and explicit expression  that only  consists of one integral  for the approximate ergodic average rate, and thus is analytically tractable.

\subsection{Jensen's Approximation for the Ergodic Achievable Rate}
It is difficult to utilize  the expression \eqref{theorem2_eq} in the optimization procedure. Therefore, we further simplify the expression of the ergodic achievable rate via the  Jensen's inequality.

\begin{theorem}
Under the narrowband SV channel model expressed in Eq. \eqref{channelG} and Eq. \eqref{channelT}, the ergodic achievable rate of the RIS-aided mmWave MIMO communication systems can be upper bounded by
\begin{equation} \label{jen1}
	\begin{aligned}
 \widetilde{R}(\mathbf{Q}, \mathbf{\Theta}) &\leq R_{Jen}(\mathbf{Q}, \mathbf{\Theta})\\
 & \triangleq \sum \limits_{i=1}^{N_s} \log _{2} \left( 1+ \frac{ N_b N_u N_r^2 }{\sigma^2 P L} d_{b,i} d_{u,i} d_{r,i}  \right).
 \end{aligned}
\end{equation}
\end{theorem}
\begin{proof}
According to the Jensen's inequality $\mathbb{E}\{\log_2 (1+x)\} \leq \log_2(1+\mathbb{E}\{x\})$ for $x\geq 0$, we have
\begin{subequations} \label{jen1_1}
\begin{align}
 \widetilde{R}(\mathbf{Q}, \mathbf{\Theta})
& =  \mathbb{E}_{g,t} \left[\sum \limits_{i=1}^{N_s} \log _{2} \left( 1+ \frac{ N_b N_u N_r^2 }{\sigma^2 P L } d_{b,i} d_{u,i} d_{r,i} |g_i|^2 |t_i|^2  \right) \right]\\
& \leq \sum \limits_{i=1}^{N_s} \log _{2} \left(1+ \mathbb{E}\left\{ \frac{ N_b N_u N_r^2 }{\sigma^2 P L } d_{b,i} d_{u,i} d_{r,i} |g_i|^2 |t_i|^2\right\} \right)\\
& = R_{Jen}(\mathbf{Q}, \mathbf{\Theta}).
\end{align}
\end{subequations}
Here, the last equality holds since $|g_i|^2 \sim \exp (1)$, $|t_i|^2 \sim \exp (1)$, and $|g_i|^2$ and $|t_i|^2$ are independent of each other.
\end{proof}

It can be found that the Jensen's approximation $R_{Jen}(\mathbf{Q}, \mathbf{\Theta})$ increases logarithmically with $N_b$, $N_u$, $N_r$ \footnote{We only consider the far-field scenario in this work.} and the eigenvalues of the matrices $\mathbf{A}_b^H {\mathbf{Q}} \mathbf{A}_b$, $\mathbf{A}_u^H \mathbf{A}_u$ and $\mathbf{X}^H \mathbf{X}$. Let us consider the special case of $\mathbf{Q}=\frac{P_T}{N_b}\mathbf{I}_{N_b}$ and $\mathbf{\Theta}=\mathbf{I}_{N_r}$. First, when ${\mathbf{Q}} = \frac{P_T}{N_b}\mathbf{I}_{N_b}$, $\{{d}_{b,i}\}_{\forall i}$ is nearly equal to $\frac{P_T}{N_b}$ because of the asymptotic orthogonality of the array response vectors when the number of antennas is large. Thus, the Jensen's approximation is independent of the number of antennas at the BS when the number of antennas exceeds a certain amount.
Second, when $\mathbf{\Theta}=\mathbf{I}_{N_r}$, we have $\mathbf{X}=\mathbf{A}_{rl}^H \mathbf{A}_{rp}$. Let us consider the asymptotic eigenvalues when the number of the reflection units at the RIS goes to infinity.  When $\phi_{r,i}^a = \psi_{r,i}^a, \phi_{r,i}^e = \psi_{r,i}^e, \forall i$, we have $\mathbf{X}\rightarrow \mathbf{I}_{L\times P}$ due to the asymptotic orthogonality of the array response vectors. Then, all non-zero eigenvalues of $\mathbf{X}^H \mathbf{X}$ is nearly equal to one, leading to the maximum value of the Jensen's approximation. When $\phi_{r,i}^a \neq \psi_{r,i}^a, \forall i$ or $\phi_{r,i}^e \neq \psi_{r,i}^e, \forall i$, we have $\mathbf{X}\rightarrow \mathbf{0}_{L\times P}$. In this case, $\{d_{r,i}\}_{\forall i}$ is nearly equal to zero, resulting in the minimum value of the Jensen's approximation. In practice, some eigenvalues of $\mathbf{X}^H \mathbf{X}$ are noticeably greater than zero, and other eigenvalues are very small due to the limited number of reflection units at the RIS. Thanks to the phase adjustment ability of RIS, the eigenvalues of $\mathbf{X}^H \mathbf{X}$ can be adjusted to be larger than zero.

\subsection{Tightness of the Approximations} \label{sec_ana_tightness}
\begin{figure}[t]
\centering
\begin{minipage}[t]{0.48\textwidth}
\centering
\includegraphics[width=8cm]{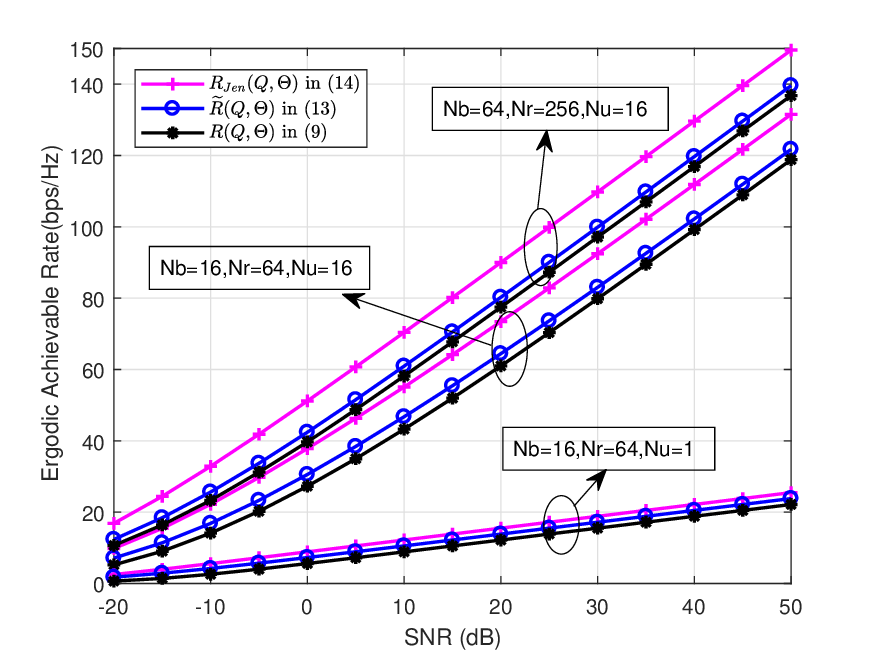}
\caption{{Ergodic achievable rate against SNR with different antenna numbers when $\mathbf{Q}=\frac{P_T}{N_b} \mathbf{I}_{N_b}$, $P=6$ and $L=8$.}} \label{cap_num}
\end{minipage}
\begin{minipage}[t]{0.48\textwidth}
\centering
\includegraphics[width=8cm]{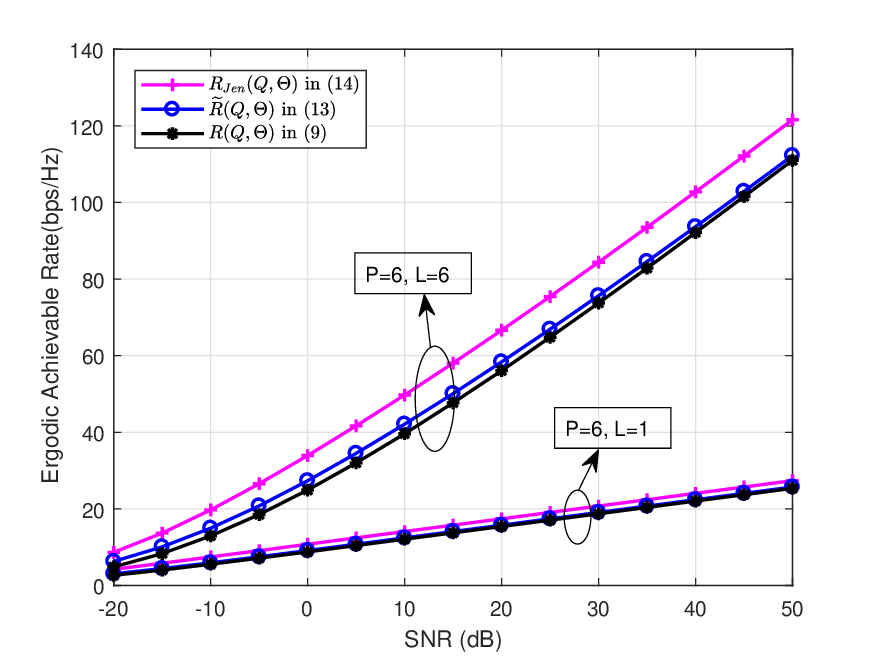}
\caption{{Ergodic achievable rate against SNR with different channel path when $\mathbf{Q}=\frac{P_T}{N_b} \mathbf{I}_{N_b}, N_b=N_u=16$ and $N_r=64$.}} \label{cap_path}
\end{minipage}
\end{figure}

\begin{figure}[t]
\centering
\begin{minipage}[t]{0.48\textwidth}
\centering
\includegraphics[width=8cm]{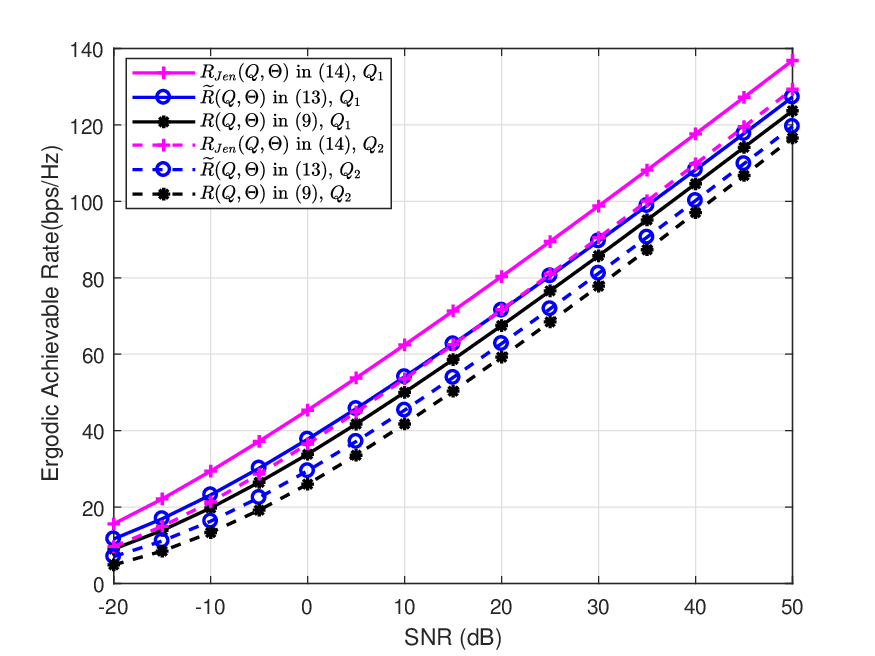}
\caption{{Ergodic achievable rate against SNR with different $\mathbf{Q}$ when $N_b=N_u=16$, $N_r=64$, $P=6$ and $L=8$.}} \label{cap_q}
\end{minipage}
\begin{minipage}[t]{0.48\textwidth}
\centering
\includegraphics[width=8cm]{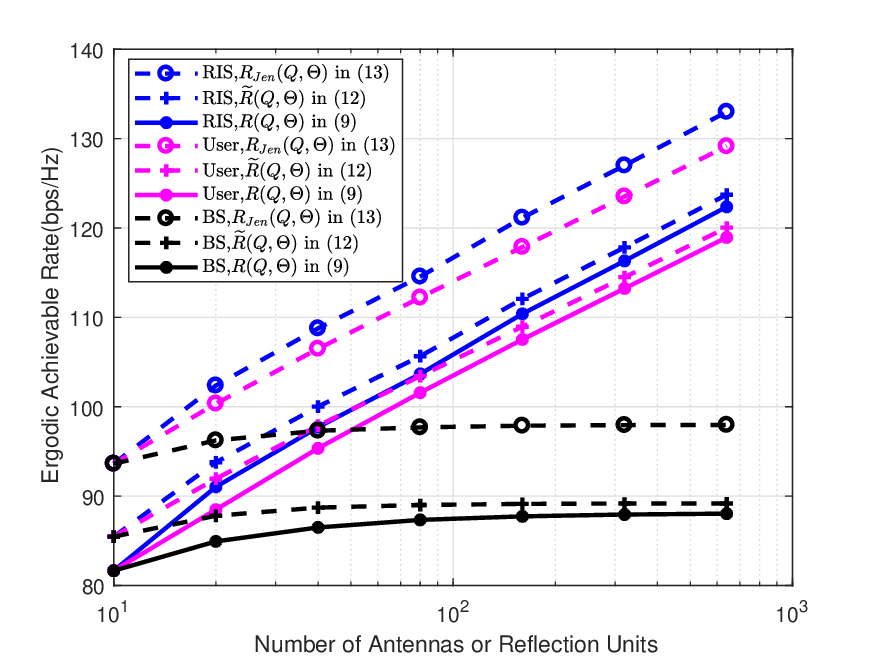}
\caption{{Ergodic achievable rate against the number of antennas or reflection units when $\mathbf{Q}=\frac{P_T}{N_b} \mathbf{I}_{N_b}$, $\text{SNR}=50$ dB and $P=L=6$.}} \label{cap_len}
\end{minipage}
\end{figure}

{In this part, we investigate the tightness of the approximations via numerical results. In the SV channel model illustrated in Section \ref{sec_2}, {the angles $\phi_{b,i}$ and $\psi_{u,i}$ are randomly generated and uniformly distributed in $[-\pi, \pi)$, $\phi^a_{r,i}$ and $\varphi^a_{r,i}$ in $[-\pi/2,\pi/2]$, and $\phi^e_{r,i}$ and $\varphi^e_{r,i}$ in $[0,\pi]$. } The response matrix $\mathbf{\Theta}$ is randomly generated where the phases $\{\theta_l\}_{\forall l}$ are uniformly and independently distributed in $[0,2\pi)$. All curves are averaged over 100 sets of independent realizations of the angles. For each set of angle realization, 1000 independent realizations of the path gains are generated for the Monte-Carlo simulations. We investigate the ergodic achievable rate against signal-to-noise ratio (SNR) with different numbers of antennas or reflection units (where $\mathbf{Q}=\frac{P_T}{N_b} \mathbf{I}_{N_b}$, $P=6$ and $L=8$), different numbers of paths (where $\mathbf{Q}=\frac{P_T}{N_b} \mathbf{I}_{N_b}$, $N_b=N_u=16$ and $N_r=8\times 8$), and different transmit covariance matrices (where $N_b=N_u=16$, $N_r=8\times 8$, $\mathbf{Q}_1= \frac{P_T}{P} \mathbf{A}_b \mathbf{I}_P \mathbf{A}_b^H$ and  $\mathbf{Q}_2= P_T \frac{\mathbf{Q}_0 \mathbf{Q}_0^H}{\operatorname{tr} \left(\mathbf{Q}_0 \mathbf{Q}_0^H\right)}$ where $\mathbf{Q}_0\sim \mathcal{C}\mathcal{N}(0,\mathbf{I}_{N_b})$)  in Fig.~\ref{cap_num},~\ref{cap_path}, and~\ref{cap_q}, respectively.
The curves of ``${R}_{Jen}(\mathbf{Q}, \mathbf{\Theta})$ in (14)" and ``$\widetilde{R}(\mathbf{Q}, \mathbf{\Theta})$ in (13)" are calculated by Eq. \eqref{jen1} and Eq. \eqref{theorem2_eq} with randomly generated angles, respectively, while the curves of ``${R}(\mathbf{Q}, \mathbf{\Theta})$ in (9)" are obtained fully by Monte-Carlo simulation. Note that, for each set of angle realization, Monte-Carlo simulation is averaged over 1000 independent realizations of channel gains. We compare the normalized error of
$R_{Jen}(\mathbf{Q}, \mathbf{\Theta})$ and $\widetilde{R}(\mathbf{Q}, \mathbf{\Theta})$ with respect to ${R}(\mathbf{Q}, \mathbf{\Theta})$ when $\text{SNR}=20$ dB in Table \ref{compu_comp}. Table \ref{compu_comp} shows that $R_{Jen}(\mathbf{Q}, \mathbf{\Theta})$ and $\widetilde{R}(\mathbf{Q}, \mathbf{\Theta})$ is about 20\% and 5\% larger than ${R}(\mathbf{Q}, \mathbf{\Theta})$ on average, respectively. From Figs.~\ref{cap_num}-\ref{cap_q} and Table \ref{compu_comp}, we find that the approximation $\widetilde{R}(\mathbf{Q}, \mathbf{\Theta})$ matches ${R}(\mathbf{Q}, \mathbf{\Theta})$ well, and the Jensen's bound $R_{Jen}(\mathbf{Q}, \mathbf{\Theta})$ is larger than ${R}(\mathbf{Q}, \mathbf{\Theta})$ with the same trend.}

\begin{table*}[!t]
\centering
\caption{Normalized error of
$R_{Jen}(\mathbf{Q}, \mathbf{\Theta})$ and $\widetilde{R}(\mathbf{Q}, \mathbf{\Theta})$ with respect to ${R}(\mathbf{Q}, \mathbf{\Theta})$ when $\text{SNR}=20$ dB.} \label{compu_comp}
\begin{tabular}{|c|c|c|c|c|c|c|c|}
\hline  \multirow{2}{*}{} & \multicolumn{3}{|c|} {\text { Fig. 2 }} & \multicolumn{2}{|c|} {\text { Fig. 3 }} & \multicolumn{2}{|c|} {\text { Fig. 4 }} \\
\cline{2-8}&  $N_b=64,$ & $N_b=16,$ & $N_b=16,$ & $P=6,$ & $P=6,$ & &\\
& $N_r=16\times 16,$ & $N_r=8 \times 8,$ & $N_r=8\times 8,$ & $L=6$ & $L=1$ &$\mathbf{Q}_1$ & $\mathbf{Q}_2$ \\
& $N_u=16$ & $N_u=16$ & $N_u=1$ & & & &\\
\hline  {$\frac{R_{Jen}(\mathbf{Q}, \mathbf{\Theta})-{R}(\mathbf{Q}, \mathbf{\Theta})}{{R}(\mathbf{Q}, \mathbf{\Theta})}$} &  16.2\% & 20.4\% & 27.3\% & 18.8\% & 13.0\% & 18.9\% & 20.8\% \\
\hline $\frac{\widetilde{R}(\mathbf{Q}, \mathbf{\Theta})-{R}(\mathbf{Q}, \mathbf{\Theta})}{{R}(\mathbf{Q}, \mathbf{\Theta})}$ &  3.6\% & 5.6\% & 13.7\%  & 4.1\% & 2.2\% & 5.9\% & 5.8\%\\
\hline
\end{tabular}
\end{table*}

{In addition, Fig.~\ref{cap_len} shows the ergodic achievable rate against the number of antennas or reflection units where $\mathbf{Q}=\frac{P_T}{N_b} \mathbf{I}_{N_b}$, $\text{SNR}=50$ dB, $P=6$ and $L=6$. The leftmost point in this figure is plotted when $N_b=N_r=N_u=10$, and the other points correspond to the case where one of the three parameters above increases to $(20, 40, 80, 160, 320, 640)$ while the other two remain at 10. Particularly, the RIS is equipped with $N_r^a\times N_r^e$ unit cells where $N_r^a=10$ and $N_r^e$ can vary.  It is found that the ergodic achievable rate increases logarithmically with the number of antennas at the user and the number of reflection units at the RIS. Furthermore, the ergodic achievable rate is nearly unchanged when the number of antennas at the BS exceeds 80. It is because the eigenvalues of $\mathbf{A}_b^H \mathbf{A}_b$ are nearly unchanged when the number of antennas at the BS is very large. Thus, the ergodic achievable rate can not be improved by increasing the number of antennas at the BS when $\mathbf{Q}=\frac{P_T}{N_b} \mathbf{I}_{N_b}$.}

\section{Optimization of Transmit Covariance Matrix and Reflection Coefficients } \label{sec_optimization}
{In Section \ref{sec_analysis}, we have derived a closed-form upper bound $R_{Jen}(\mathbf{Q}, \mathbf{\Theta})$  of $R(\mathbf{Q}, \mathbf{\Theta})$.
Therefore, we replace $R(\mathbf{Q}, \mathbf{\Theta})$ in $\mathcal{P}_0$ with $R_{Jen}(\mathbf{Q}, \mathbf{\Theta})$, yielding}
    \begin{subequations}\label{prob_app}
    \begin{align}
            \mathcal{P}_1: \max \limits_{ \mathbf{Q}, \mathbf{\Theta} } \quad & \sum \limits_{i=1}^{N_s} \log _{2} \left( 1+  \frac{ N_b N_u N_r^2 }{\sigma^2 P L} d_{b,i} d_{u,i} d_{r,i} \right) \\
            \text { s.t. } \quad & \eqref{const_q1},\eqref{const_q2},\eqref{const_ris},
    \end{align}
    \end{subequations}
where $d_{b,i}$, $d_{u,i}$, and $d_{r,i}$ are defined in \textbf{Lemma} \ref{lemma1}. Compared with the original problem $\mathcal{P}_0$, the problem $\mathcal{P}_1$ is much clearer. However, the problem $\mathcal{P}_1$ is still challenging due to two facts. Firstly, the objective function is not directly related to the optimizing variables $\mathbf{Q}$ and $\mathbf{\Theta}$. Thus, we need to bridge the gap for the further optimization procedure. Secondly, the problem $\mathcal{P}_1$ is highly non-convex due to the unit-modulus constraints. In the following, we will first introduce a lemma to connect the objective function with the optimizing variables. Then, the alternating optimization method is adopted to decouple the problem $\mathcal{P}_1$ into two sub-problems.

\begin{lemma} \label{theorem_opt}
Let $\mathbf{U}\in\mathbb{C}^{N\times N}$ be a unitary matrix, i.e., $\mathbf{U}^H \mathbf{U} = \mathbf{U} \mathbf{U}^H= \mathbf{I}$. If $\mathbf{c} \in \mathbb{R}_+^N$ and $\mathbf{s} \in \mathbb{R}_+^N$ are nonnegative real sequences that are in descending order. Then, we have
\begin{equation}
\begin{aligned}	
  &\operatorname{det} \left[ \mathbf{I} + \operatorname{diag}(\mathbf{c}) \mathbf{U}^H \operatorname{diag}(\mathbf{s}) \mathbf{U}\right]
 \leq \\
 &\operatorname{det}   \left[ \mathbf{I} + \mathbf{U}^H  \operatorname{diag}(\mathbf{c}) \operatorname{diag}(\mathbf{s}) \mathbf{U}\right]
\end{aligned} 
\end{equation}
\end{lemma}
\begin{proof}
See Appendix \ref{append2}.
\end{proof}

\subsection{Transmit Covariance Matrix Optimization Given RIS Reflection Coefficients}
Define $\beta =  \frac{ N_b N_u N_r^2 }{\sigma^2 P L} $ and $ \boldsymbol{\gamma} = \beta \operatorname{diag} \left(d_{u,1} d_{r,1}, d_{u,2} d_{r,2}, \ldots, d_{u,N_s} d_{r,N_s},0,\ldots,0\right)\in\mathbb{R}^{P\times P}$, then we have
\begin{subequations}
\begin{align}
R_{Jen}(\mathbf{Q}, \mathbf{\Theta}) &= \sum \limits_{i=1}^{N_s} \log _{2} \left( 1+ \beta d_{b,i} d_{u,i} d_{r,i}  \right) \\
& = \log _2 \operatorname{det}\left[ \mathbf{I}_{P} + \boldsymbol{\gamma} \mathbf{D}_b \right] \\
& \overset{(a)}{\geq} \log _2 \operatorname{det}\left[ \mathbf{I}_{P} + \boldsymbol{\gamma} \mathbf{U}_b^H \mathbf{D}_b \mathbf{U}_b \right]\\
& \overset{(b)}{=} \log _2 \operatorname{det}\left[ \mathbf{I}_{P} + \boldsymbol{\gamma} \mathbf{A}_b^H \mathbf{Q} \mathbf{A}_b \right] \\
& = \log _2 \operatorname{det}\left[ \mathbf{I}_{P} + \left(\mathbf{A}_b \boldsymbol{\gamma}^{1/2}\right)^H  \mathbf{Q} \left(\mathbf{A}_b \boldsymbol{\gamma}^{1/2}\right) \right],
\end{align}
\end{subequations}
where $(a)$ holds due to \textbf{Lemma} \ref{theorem_opt} and $(b)$ holds due to $\mathbf{A}_b^H \mathbf{Q} \mathbf{A}_b = \mathbf{U}_{b}^H \mathbf{D}_b \mathbf{U}_{b}$. Note that we find the lower bound of the Jensen's approximation through \textbf{Lemma} \ref{theorem_opt}. Thus, it is reasonable to maximize $R_{Jen}(\mathbf{Q}, \mathbf{\Theta})$ by maximizing its lower bound. Therefore, when fixing the reflection coefficients at the RIS, the problem $\mathcal{P}_1$ can be transformed into
    \begin{subequations}\label{prob_Q}
    \begin{align}
             \max \limits_{ \mathbf{Q} } \quad & \log _2 \operatorname{det}\left[ \mathbf{I}_{P} + \left(\mathbf{A}_b \boldsymbol{\gamma}^{1/2}\right)^H  \mathbf{Q} \left(\mathbf{A}_b \boldsymbol{\gamma}^{1/2}\right) \right] \\
            \text { s.t. } \quad & \operatorname{tr}(\mathbf{Q}) \leq P_T, \\
            & \mathbf{Q} \succeq 0.
    \end{align}
    \end{subequations}
Note that it is  a convex optimization problem over $\mathbf{Q}$ and can be optimally solved by the well-known water-filling algorithm. Specifically, define the singular value decomposition (SVD) of $ \left(\mathbf{A}_b \boldsymbol{\gamma}^{1/2}\right)^H$ as
\begin{equation}
  \left(\mathbf{A}_b \boldsymbol{\gamma}^{1/2}\right)^H =\mathbf{U}\boldsymbol{\Sigma} \mathbf{V}^H = \left[\begin{array}{ll}
\boldsymbol{U}_{1} & \boldsymbol{U}_{2}
\end{array}\right]\left[\begin{array}{cc}
\boldsymbol{\Sigma}_{1} & \boldsymbol{0} \\
\boldsymbol{0} & \boldsymbol{\Sigma}_{2}
\end{array}\right]\left[\begin{array}{l}
\boldsymbol{V}_{1}^{H} \\
\boldsymbol{V}_{2}^{H}
\end{array}\right],
\end{equation}
where $\mathbf{U}$ is a $P \times P$ unitary matrix, $\boldsymbol{\Sigma}$ is a $P \times N_b$ diagonal matrix of singular values, $\mathbf{V}$ is an $N_b \times N_b$ unitary matrix. The truncated SVD of $ \left(\mathbf{A}_b \boldsymbol{\gamma}^{1/2}\right)^H$ is $\mathbf{U}_1 \boldsymbol{\Sigma}_1 \mathbf{V}_1^H$, where $\boldsymbol{\Sigma}_1 \in \mathbb{R}^{P\times N_s}$ and $\mathbf{V}_1 \in \mathbb{C}^{N_b \times N_s}$. Thus, the optimal solution to the problem \eqref{prob_Q} is given by
\begin{equation} \label{opt_Q}
  \mathbf{Q}^\star = \mathbf{V}_1 \operatorname{diag}(p_1, p_2, \ldots, p_{N_s} ) \mathbf{V}_1^H,
\end{equation}
where $p_i = \max \left( 1/p_0 -1/\Sigma_1^2(i,i), 0 \right), i=1,2, \ldots, N_s$ denotes the optimal amount of power allocated to the $i$-th data stream, and $p_0$ is the water level satisfying $\sum_{i=1}^{N_s} p_i = P_T$.

\subsection{Reflection Coefficients Optimization Given Transmit Covariance Matrix}
Define $\boldsymbol{\Gamma} = \beta \operatorname{diag} \left(d_{u,1} d_{b,1}, d_{u,2} d_{b,2}, \ldots, d_{u,N_s} d_{b,N_s}, 0,\ldots,0\right)\in\mathbb{R}^{P\times P}$, similar to the transmit covariance matrix optimization, we have
\begin{subequations}
\begin{align}
R_{Jen}(\mathbf{Q}, \mathbf{\Theta}) &= \sum \limits_{i=1}^{N_s} \log _{2} \left( 1+ \beta d_{b,i} d_{u,i} d_{r,i}  \right) \\
& = \log _2 \operatorname{det}\left[ \mathbf{I}_{P} + \boldsymbol{\Gamma} \mathbf{D}_r \right] \\
& \overset{(a)}{\geq} \log _2 \operatorname{det}\left[ \mathbf{I}_{P} + \boldsymbol{\Gamma} \mathbf{U}_r^H \mathbf{D}_r \mathbf{U}_r \right]\\
& \overset{(b)}{=} \log _2 \operatorname{det}\left[ \mathbf{I}_{P} + \boldsymbol{\Gamma} \mathbf{X}^H \mathbf{X} \right],
\end{align}
\end{subequations}
where $(a)$ holds due to \textbf{Lemma} \ref{theorem_opt}, $(b)$ holds due to $\mathbf{X}^H \mathbf{X} = \mathbf{U}_{r}^H \mathbf{D}_r \mathbf{U}_{r}$ and $\mathbf{X} = \mathbf{A}_{rl}^H \mathbf{\Theta} \mathbf{A}_{rp}$. Therefore, when fixing the transmit covariance matrix at the BS, the problem $\mathcal{P}_1$ is reduced to
\begin{subequations}\label{prob_ris}
\begin{align}
\max \limits_{ \mathbf{\Theta} } \quad& \log _2 \operatorname{det}\left[ \mathbf{I}_{P} + \boldsymbol{\Gamma} \mathbf{A}_{rp}^H \mathbf{\Theta}^H \mathbf{A}_{rl} \mathbf{A}_{rl}^H \mathbf{\Theta} \mathbf{A}_{rp} \right] \\
\text { s.t. } \quad & \mathbf{\Theta} =\operatorname{diag}\left(e^{j \phi_{1}}, \cdots, e^{j \phi_{N_r}} \right).
\end{align}
\end{subequations}
The problem \eqref{prob_ris} is highly non-convex due to the non-convex unit-modulus constraints and the fact that the objective function is not concave with respect to $\mathbf{\Theta}$. Here, we adopt the RCG algorithm to handle it. The RCG algorithm is widely applied in hybrid beamforming design \cite{7397861} and recently applied in RIS-aided systems as well \cite{8982186,9234098, 8855810, 9090356}. Note that the RCG algorithm is suitable to handle the problems with the unit-modulus constraints, and it usually can carry out a good solution.

Note that the reflection coefficients of the RIS have a diagonal structure. Thus, we introduce a mask matrix $\mathbf{I}$ ($\mathbf{I}$ is an identity matrix). Then, we can rewrite the reflection coefficients of the RIS as $\mathbf{\Theta}=\mathbf{I} \odot \widetilde{\mathbf{\Theta}}$, where $\widetilde{\mathbf{\Theta}}\in \mathbb{C}^{N_r \times N_r}$ is an auxiliary matrix variable without the diagonal constraint and all of its units satisfy the unit-modulus constraints. Thereby, the feasible set of $\widetilde{\mathbf{\Theta}}$ forms a Riemannian manifold $\mathcal{M}= \{\widetilde{\mathbf{\Theta}}\in \mathbb{C}^{N_r \times N_r}: |\widetilde{\mathbf{\Theta}}_{ij}|=1, \forall i,j\}$ \cite{AbsilMahonySepulchre+2009}.  {The main idea of the RCG algorithm is to generalize a conjugate gradient method from the Euclidean space to the manifold space, with the main challenge of calculating the Euclidean gradient.} Here, we start with the definition of the tangent space. The tangent space ${T}_{\widetilde{\mathbf{\Theta}}_i} \mathcal{M}$ of the manifold $\mathcal{M}$ at point $\widetilde{\mathbf{\Theta}}_i$ is given by
\begin{equation}
{T}_{\widetilde{\mathbf{\Theta}}_i} \mathcal{M} =\left\{\mathbf{Z} \in \mathbb{C}^{N_r \times N_r}: \Re\left\{\mathbf{Z} \odot \widetilde{\mathbf{\Theta}}_i^{*}\right\}=\mathbf{0}\right\}.
\end{equation}
The main procedure of the RCG algorithm consists of three steps in each iteration.

1) \textsl{Riemannian Gradient}: The Riemannian gradient is one tangent vector (direction) with the steepest increase of the objective function. Define $\mathbf{\Omega}= \mathbf{I}_{P} + \boldsymbol{\Gamma} \mathbf{A}_{rp}^H \mathbf{\Theta}^H \mathbf{A}_{rl} \mathbf{A}_{rl}^H \mathbf{\Theta} \mathbf{A}_{rp}$, the Riemannian gradient of the objective function $f(\widetilde{\mathbf{\Theta}})= -\log_2 \operatorname{det}(\mathbf{\Omega})$ at the point $\widetilde{\mathbf{\Theta}}_i$, denoted by $\operatorname{grad} f(\widetilde{\mathbf{\Theta}}_i) $,  is given by
\begin{equation} \label{grad}
\operatorname{grad} f(\widetilde{\mathbf{\Theta}}_i)=\nabla f(\widetilde{\mathbf{\Theta}}_i) -\Re \left\{\nabla f(\widetilde{\mathbf{\Theta}}_i) \odot \widetilde{\mathbf{\Theta}}_i^*\right\} \odot \widetilde{\mathbf{\Theta}}_i,
\end{equation}
where $\nabla f(\widetilde{\mathbf{\Theta}}_i)$ denotes the Euclidean gradient. In the following, we will compute the Euclidean gradient of the objective function.

Based on the differential rule $\operatorname{d}(\operatorname{det}(\mathbf{A})) = \operatorname{det} (\mathbf{A}) \operatorname{tr}(\mathbf{A}^{-1}  \operatorname{d}(\mathbf{A}))$, we have
\begin{subequations}
\begin{align}
\operatorname{d} \left( f(\widetilde{\mathbf{\Theta}}) \right)
&=-\operatorname{d} \left(\log_2 \operatorname{det}(\mathbf{\Omega})\right)\\
&= -\frac{1}{\ln 2} \operatorname{tr} (\mathbf{\Omega}^{-1} \operatorname{d} (\mathbf{\Omega}) )\\
&= -\frac{1}{\ln 2} \operatorname{tr} (\mathbf{\Omega}^{-1} \boldsymbol{\Gamma} \mathbf{A}_{rp}^H \operatorname{d}(\mathbf{\Theta}^H) \mathbf{A}_{rl} \mathbf{A}_{rl}^H \mathbf{\Theta} \mathbf{A}_{rp}  ) \\
&\overset{(a)}{=} -\frac{1}{\ln 2}\operatorname{tr} ( \mathbf{A}_{rl} \mathbf{A}_{rl}^H \mathbf{\Theta} \mathbf{A}_{rp} \mathbf{\Omega}^{-1} \boldsymbol{\Gamma} \mathbf{A}_{rp}^H \operatorname{d}(\mathbf{\Theta}^H) ) \\
& \overset{(b)}{=} -\frac{1}{\ln 2}\operatorname{tr} \Big( \left(\mathbf{A}_{rl} \mathbf{A}_{rl}^H \mathbf{\Theta} \mathbf{A}_{rp} \mathbf{\Omega}^{-1} \boldsymbol{\Gamma} \mathbf{A}_{rp}^H \right)  \notag \\
&\quad \;  \times \left( \mathbf{I} \odot \operatorname{d}\left(\widetilde{\mathbf{\Theta}}^H\right)\right) \Big)  \\
& \overset{(c)}{=} -\frac{1}{\ln 2}\operatorname{tr} \Big( \big( \left(\mathbf{A}_{rl} \mathbf{A}_{rl}^H \mathbf{\Theta} \mathbf{A}_{rp} \mathbf{\Omega}^{-1} \boldsymbol{\Gamma} \mathbf{A}_{rp}^H \right)  \notag \\
&\quad \;  \odot \mathbf{I} \big) \operatorname{d}\left(\widetilde{\mathbf{\Theta}}^H\right) \Big),
\end{align}
\end{subequations}
where $(a)$ holds due to $\operatorname{tr}(\mathbf{AB})= \operatorname{tr} (\mathbf{BA})$ for arbitrary matrices $\mathbf{A}$ and $\mathbf{B}$, $(b)$ holds due to $\operatorname{d}(\mathbf{\Theta}) =  \mathbf{I} \odot \operatorname{d}(\widetilde{\mathbf{\Theta}})$, and $(c)$ holds due to $ \operatorname{tr}(\mathbf{A} (\mathbf{B} \odot \mathbf{C} )) = \operatorname{tr} ( \left(\mathbf{A} \odot \mathbf{B}^T \right) \mathbf{C} )$ for arbitrary matrices $\mathbf{A}$, $\mathbf{B}$ and $\mathbf{C}$.

According to the fact that $\operatorname{d} \left( f(\widetilde{\mathbf{\Theta}}) \right) = \operatorname{tr} \left( \nabla  f(\widetilde{\mathbf{\Theta}})   \operatorname{d}(\widetilde{\mathbf{\Theta}}^H) \right) $, we obtain the Euclidean gradient of the objective function $f(\widetilde{\mathbf{\Theta}})$ as
\begin{equation} \label{grad_eu}
  \nabla  f(\widetilde{\mathbf{\Theta}}) = -\frac{1}{\ln 2} \left(\mathbf{A}_{rl} \mathbf{A}_{rl}^H \mathbf{\Theta} \mathbf{A}_{rp} \mathbf{\Omega}^{-1} \boldsymbol{\Gamma} \mathbf{A}_{rp}^H \right) \odot \mathbf{I}.
\end{equation}

2) \textsl{Transport}: With the obtained Riemannian gradient, the optimization tools in the Euclidean space can be extended to the manifold space. However, the search direction $\boldsymbol{\eta}_i$ and $\boldsymbol{\eta}_{i+1}$ usually lie in two different tangent spaces. Therefore, the transport operation ${T}_{\widetilde{\mathbf{\Theta}}_i \rightarrow \mathbf{\widetilde{\mathbf{\Theta}}_{i+1}}} \left(\boldsymbol{\eta}_{i}\right)$ is needed to map the tangent vector $\boldsymbol{\eta}_i$ from ${T}_{\widetilde{\mathbf{\Theta}}_i} \mathcal{M}$ to ${T}_{\widetilde{\mathbf{\Theta}}_i+1} \mathcal{M}$, which is given by
\begin{equation} \label{transport}
\begin{aligned}
\mathcal{T}_{\widetilde{\mathbf{\Theta}}_i \rightarrow \widetilde{\mathbf{\Theta}}_{i+1}}\left(\boldsymbol{\eta}_{i}\right) : T_{\widetilde{\mathbf{\Theta}}_i} \mathcal{M} & \mapsto T_{\widetilde{\mathbf{\Theta}}_{i+1}} \mathcal{M}: \\
\boldsymbol{\eta}_{i} & \mapsto \boldsymbol{\eta}_{i}-\Re\left\{\boldsymbol{\eta}_{i} \odot \widetilde{\mathbf{\Theta}}_{i+1}^{*}\right\} \odot \widetilde{\mathbf{\Theta}}_{i+1}.
\end{aligned}
\end{equation}
Then, we can update the search direction as
\begin{equation}\label{direction}
\boldsymbol{\eta}_{i+1}=-\operatorname{grad} f(\widetilde{\mathbf{\Theta}}_{i+1}) + \beta_i \mathcal{T}_{\widetilde{\mathbf{\Theta}}_i \rightarrow \widetilde{\mathbf{\Theta}}_{i+1}}\left(\boldsymbol{\eta}_{i}\right),
\end{equation}
where $\beta_i$ is chosen as the Polak-Ribiere parameter \cite{AbsilMahonySepulchre+2009}.

3) \textsl{Retraction}: After determining the search direction $\eta_i$, we need to determine the step size $\alpha_i$. However, the obtained point $\alpha_i \boldsymbol{\eta}_i$ may leave the manifold. Thereby, an operation called \textsl{retraction} is needed to map it from the tangent space to the manifold itself, which is given by
\begin{equation} \label{retraction}
\begin{aligned}
\mathcal{R}_{\widetilde{\mathbf{\Theta}}_i}(\alpha_i \boldsymbol{\eta}_i) : T_{\widetilde{\mathbf{\Theta}}_i} \mathcal{M} & \mapsto \mathcal{M}: \\
\alpha_i \boldsymbol{\eta}_{i} & \mapsto \frac{\left(\widetilde{\mathbf{\Theta}}_i + \alpha_i \boldsymbol{\eta}_{i}\right)_{j}}{\left| \left(\widetilde {\mathbf{\Theta}}_i + \alpha_i \boldsymbol{\eta}_{i}\right)_{j}\right|},
\end{aligned}
\end{equation}
where $(\widetilde{\mathbf{\Theta}}_i+ \alpha_i \boldsymbol{\eta}_{i})_j$ denotes the $j$-th entry of $\widetilde{\mathbf{\Theta}}_i+ \alpha_i \boldsymbol{\eta}_{i}$.

\begin{algorithm}[t]
    \caption{RCG Algorithm for Reflection Coefficients Optimization}
    \label{rcg}
    \begin{algorithmic}[1]
    \State {\bf Input:} $\{\mathbf{\Gamma}, \mathbf{A}_{rp}, \mathbf{A}_{rl}\}$, desired accuracy $\epsilon$
    \State Initialize: $\widetilde{\mathbf{\Theta}}_0$, $\boldsymbol{\eta}_0= -\operatorname{grad} f(\widetilde{\mathbf{\Theta}}_0)$, and set $i=0$;
    \Repeat
            \State Choose the Armijo backtracking line search step size $\alpha_i$;
            \State  Find the next point $\widetilde{\mathbf{\Theta}}_{i+1}$ using retraction in \eqref{retraction}: $\widetilde{\mathbf{\Theta}}_{i+1} = \mathcal{R}_{\widetilde{\mathbf{\Theta}}_i}(\alpha_i \boldsymbol{\eta}_i)$;
            \State Calculate the Euclidean gradient $\nabla  f(\widetilde{\mathbf{\Theta}}_{i+1})$ according to \eqref{grad_eu};
            \State Calculate the Riemannian gradient $\operatorname{grad} f(\widetilde{ \mathbf{\Theta}}_{i+1})$ according to \eqref{grad};
            \State Calculate the transport $\mathcal{T}_{\widetilde{\mathbf{\Theta}}_i \rightarrow \widetilde{\mathbf{\Theta}}_{i+1}} \left(\boldsymbol{\eta}_{i}\right)$ according to \eqref{transport};
            \State Calculate the conjugate direction $\boldsymbol{\eta}_{i+1}$ according to \eqref{direction};
            \State $i \leftarrow i+1$;
    \Until $\|\operatorname{grad} f(\widetilde{\mathbf{\Theta}}_i) \|_2 \leq \epsilon$.
    \State {\bf Output:} $\widetilde{\mathbf{\Theta}}^\star = \widetilde{\mathbf{\Theta}}_i$, ${\mathbf{\Theta}}^\star = \mathbf{I} \odot \widetilde{\mathbf{\Theta}}^\star$.
    \end{algorithmic}
 \end{algorithm}
The key steps used in each iteration of the manifold optimization are introduced above, and the consequent algorithm for reflection coefficients optimization is summarized in \textbf{Algorithm} \ref{rcg}. \textbf{Algorithm} \ref{rcg} is guaranteed to converge to a stationary point \cite{AbsilMahonySepulchre+2009}.

\subsection{Overall Algorithm}
In the above two subsections, the transmit covariance matrix and the reflection coefficients are optimized in each sub-problem. Here, we propose the overall alternating optimization algorithm for problem $\mathcal{P}_1$ in \textbf{Algorithm} \ref{alg_overall}. Firstly, the transmit covariance matrix $\mathbf{Q}^{(0)}$ is initialized as an identity matrix, and the reflection coefficients $\mathbf{\Theta}^{(1)}$ is obtained based on \textbf{Algorithm} \ref{rcg}. Then, the transmit covariance matrix $\mathbf{Q}^{(1)}$ is updated according to \eqref{opt_Q} with fixed $\mathbf{\Theta}^{(1)}$. In each iteration, the lower bound of the objective value of problem $\mathcal{P}_1$ is maximized. Thus, the objective value of problem $\mathcal{P}_1$ is non-decreasing over iterations. Therefore, by iteratively calculating $\mathbf{Q}^{(i+1)}$ and $\mathbf{\Theta}^{(i+1)}$, our proposed algorithm is guaranteed to converge.

Now, let us consider the complexity of the proposed \textbf{Algorithm} \ref{alg_overall}. Firstly, the complexity of the water-filling algorithm is $\mathcal{O} (PN_b \min(P,N_b)) = \mathcal{O}(N_b)$. Secondly, the complexity of the RCG algorithm is dominated by calculating the Euclidean gradient \eqref{grad_eu}. Define $\Xi_1 = \mathbf{A}_{rl}^H \mathbf{\Theta} \mathbf{A}_{rp} \mathbf{\Omega}^{-1} \boldsymbol{\Gamma}$ and $\Xi_2 = \mathbf{A}_{rl} \Xi_1 \mathbf{A}_{rp}^H$. Note that only the diagonal elements of $\Xi_2$ need to be calculated, and the $i$-th diagonal element of $\Xi_2$ can be calculated by multiplying the $i$-th row of $\mathbf{A}_{rl}$, $\Xi_1$ and the $i$-th column of $\mathbf{A}_{rp}^H$. Thus, the complexity of \textbf{Algorithm} \ref{rcg} is $\mathcal{O}(N_r I_1)$, where $I_1$ denotes the number of iterations of the RCG algorithm. Therefore, the overall complexity is $\mathcal{O} ( (N_b + N_r I_1) I_2)$, where $I_2$ denotes the number of iterations of \textbf{Algorithm} \ref{alg_overall}.

\begin{algorithm}[t]
    \caption{Alternating Optimization Algorithm for Problem $\mathcal{P}_1$}
    \label{alg_overall}
    \begin{algorithmic}[1]
    \State Initialize: $\mathbf{Q}^{(0)}= \frac{P_T}{N_b} \mathbf{I}_{N_b}$, $\mathbf{\Theta}^{(0)}$ is randomly generated where the phases $\{\theta_l\}_{\forall l}$ are uniformly and independently distributed in $[0,2\pi)$, error tolerance $\epsilon$, and set $i=0$;
    \Repeat
        \State Calculate $\mathbf{\Theta}^{(i+1)}$ based on \textbf{Algorithm} \ref{rcg} with fixed $\mathbf{Q}^{(i)}$;
        \State Calculate $\mathbf{Q}^{(i+1)}$ according to \eqref{opt_Q} with fixed $\mathbf{\Theta}^{(i+1)}$;
        \State $i \leftarrow i+1$;
    \Until The increase of the objective value of the problem $\mathcal{P}_1$ is below the threshold $\epsilon$.
    \end{algorithmic}
 \end{algorithm}

\section{Simulation Results} \label{sec_simulation}
In this section, we evaluate the performance of the proposed algorithms for ergodic achievable rate optimization. As in Section \ref{sec_ana_tightness}, {the angles $\phi_{b,i}$ and $\psi_{u,i}$ are randomly generated and uniformly distributed in $[-\pi, \pi)$, $\phi^a_{r,i}$ and $\varphi^a_{r,i}$ in $[-\pi/2,\pi/2]$, and $\phi^e_{r,i}$ and $\varphi^e_{r,i}$ in $[0,\pi]$. } Other system parameters are set as follows unless specified otherwise later: $N_b=16, N_r=8\times 8, N_u =16, P=6, L=8$. All curves are averaged over 100 sets of independent realizations of the angles. For each set of angle realization, 1000 independent realizations of the path gains are generated for the Monte-Carlo simulations.

\begin{figure}[t]
\centering
\begin{minipage}[t]{0.48\textwidth}
\centering
\includegraphics[width=8cm]{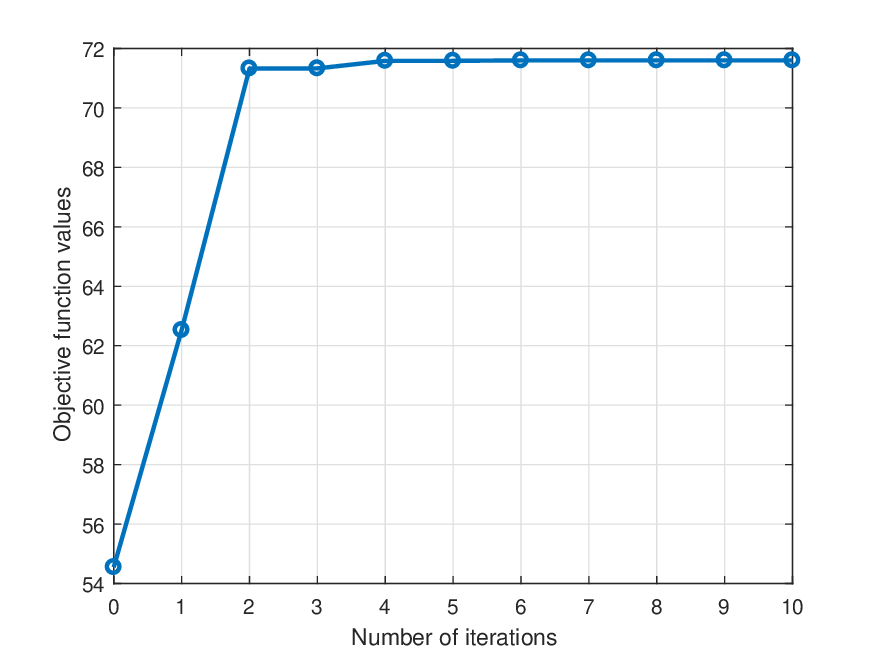}
 \caption{{Convergence of the proposed algorithm with $\text{SNR}=10$ dB.}} \label{conver}
\end{minipage}
\begin{minipage}[t]{0.48\textwidth}
\centering
\includegraphics[width=8cm]{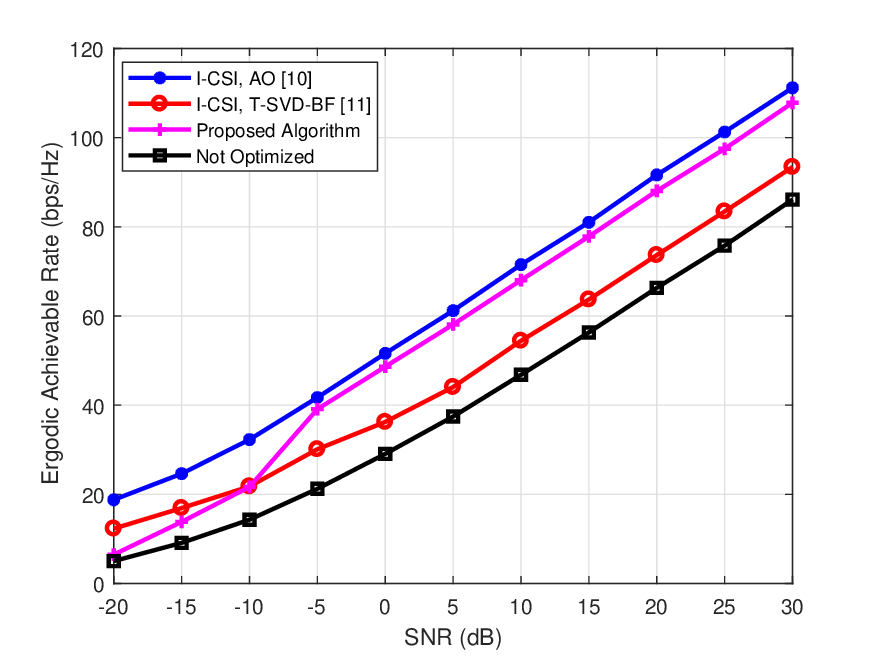}
 \caption{{Performance comparison with benchmarks. }} \label{bench_comp}
\end{minipage}
\end{figure}

Fig.~\ref{conver} illustrates the convergence performance of the proposed \textbf{Algorithm} \ref{alg_overall}. It can be found that the proposed algorithm converges in just four iterations, which verifies the convergence of the proposed algorithm.

In Fig.~\ref{bench_comp}, we compare the rate performance achieved by the proposed algorithm with statistical CSI against two state-of-the-art benchmarks that assume the knowledge of perfect instantaneous CSI (I-CSI).
\begin{itemize}
  \item Alternating optimization based algorithm (AO) \cite{9110912}: The transmit covariance matrix $\mathbf{Q}$ and each unit of the reflection coefficients of the RIS $\{\theta_i\}_{\forall i}$ are alternately optimized until convergence. Note that the closed-form solution at each step is obtained.
  \item Truncated-SVD-based beamforming (T-SVD-BF) \cite{9234098}: By exploiting the sparse structure of the mmWave communication, work \cite{9234098} first proposes a manifold-based method to optimize the reflection coefficients of the RIS, then adopts the water-filling algorithm to handle the active beamforming at the BS. Note that no alternating process is needed and each step only needs to be executed once.
\end{itemize}
Note that the curve ``Not Optimized'' indicates that $\mathbf{Q} = \frac{P_T}{N_b} \mathbf{I}_{N_b}$ and $\mathbf{\Theta}$ is randomly generated by setting  the phases $\{\theta_l\}_{\forall l}$ as uniformly and independently distributed in $[0,2\pi)$. The curve ``Proposed Algorithm'' is obtained through the following procedure: $\mathbf{Q}$ and $\mathbf{\Theta}$ are first obtained by \textbf{Algorithm} \ref{alg_overall}, based on which Monte-Carlo simulations are performed to get the corresponding ergodic achievable rate. It is found that our proposed algorithm outperforms T-SVD-BF. When SNR is low, our proposed algorithm performs much worse than the AO-based algorithm. Nevertheless, when SNR is larger than -5 dB, the proposed algorithm approaches the AO-based algorithm, with about 3 bits/s/Hz performance loss. The AO-based algorithm performs best because it has closed-form solution in each step and converges to a stationary point. However, when the number of reflection units at the RIS is large, its computational complexity is very high, which is not suitable in practice. The T-SVD-BF method exploits the sparse structure of the mmWave channel and runs very fast. However, its performance is the worst. Note that both the AO-based algorithm and the T-SVD-BF method need perfect instantaneous CSI, which is difficult to obtain. Nevertheless, our proposed algorithm only needs the statistical CSI and its performance is attractive.
\begin{figure}[t]
\centering
\begin{minipage}[t]{0.48\textwidth}
\centering
\includegraphics[width=8cm]{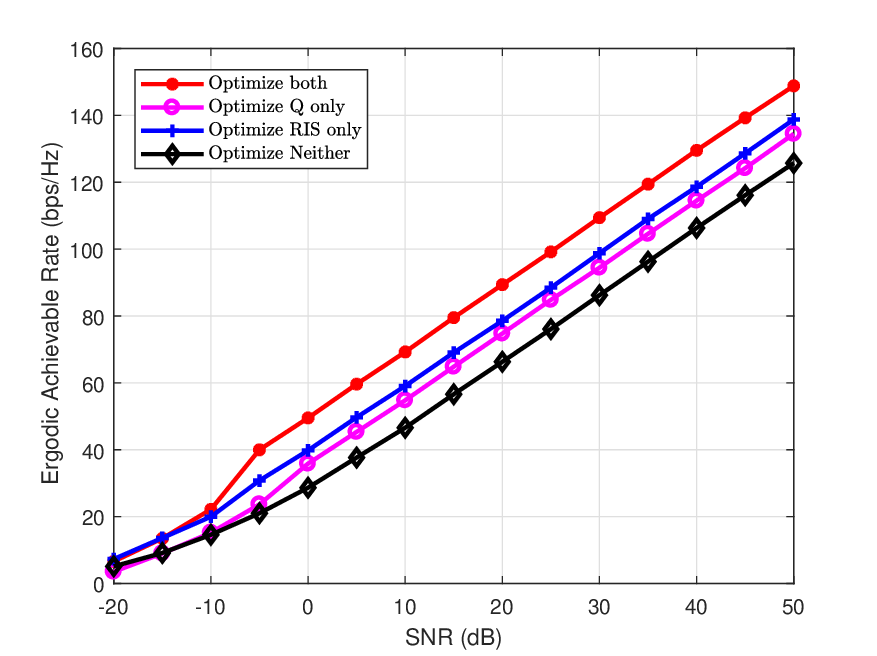}
 \caption{{Influence comparison between transmit covariance matrix and reflection coefficients  with $N_b=16,N_r=64$ and $N_u=16$.}} \label{opt_Q_ris1}
\end{minipage}
\begin{minipage}[t]{0.48\textwidth}
\centering
\includegraphics[width=8cm]{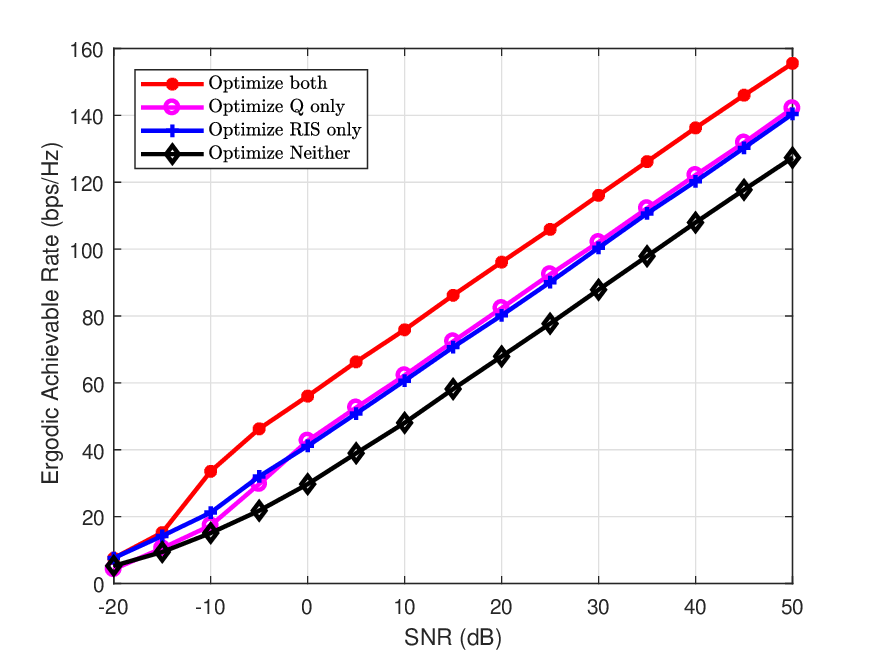}
 \caption{{Influence comparison between transmit covariance matrix and reflection coefficients  with $N_b=32,N_r=64$ and $N_u=16$.}} \label{opt_Q_ris2}
\end{minipage}
\end{figure}

\begin{figure}[t]
\centering
\begin{minipage}[t]{0.48\textwidth}
\centering
\includegraphics[width=8cm]{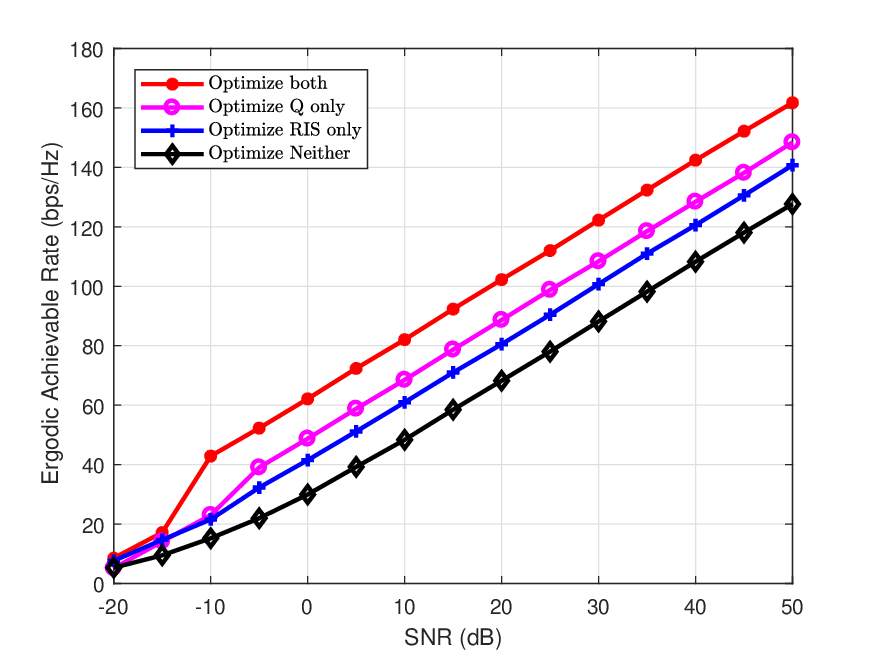}
 \caption{{Influence comparison between transmit covariance matrix and reflection coefficients  with $N_b=64,N_r=64$ and $N_u=16$.}} \label{opt_Q_ris3}
\end{minipage}
\begin{minipage}[t]{0.48\textwidth}
\centering
\includegraphics[width=8cm]{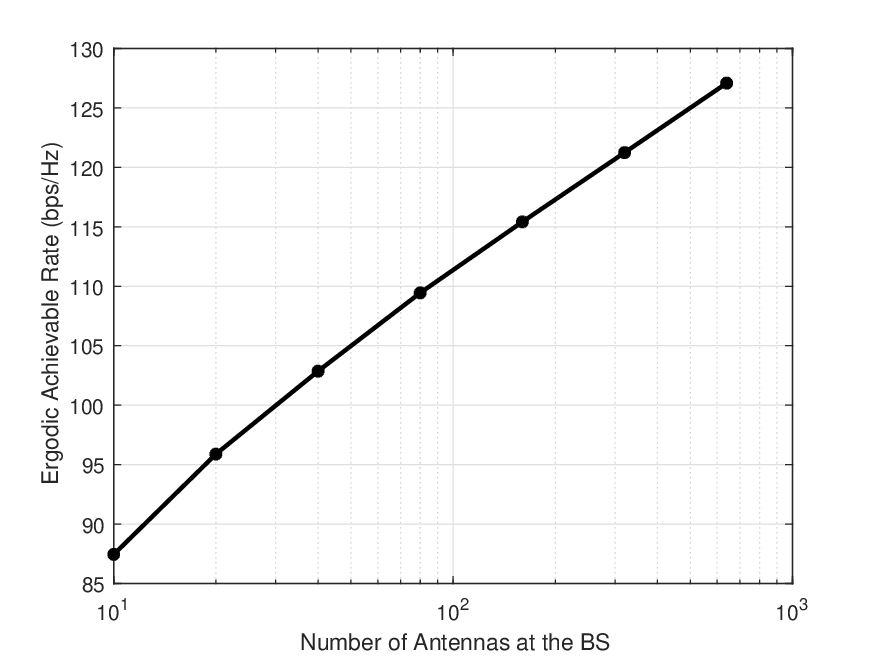}
 \caption{{Ergodic achievable rate against the number of antennas or reflection units with $\text{SNR}=50$ dB, $P=L=6$.}} \label{main_rislength}
\end{minipage}
\end{figure}

In addition, we compare the influence on ergodic achievable rate of transmit covariance matrix and reflection coefficients in Fig.~\ref{opt_Q_ris1} - \ref{opt_Q_ris3}, which are obtained similar to the curve ``Proposed Algorithm'' in Fig. \ref{bench_comp}.  The only difference among them is the number of antennas at the BS, which changes from 16, 32 to 64. The curve ``Optimize RIS only" indicates that $\mathbf{Q} = \frac{P_T}{N_b} \mathbf{I}_{N_b}$. The curve ``Optimize Q only" indicates that $\mathbf{\Theta}$ is randomly generated by setting  the phases $\{\theta_l\}_{\forall l}$ as uniformly and independently distributed in $[0,2\pi)$. Firstly, it is found that the ergodic achievable rate after optimization is about 20 bits/s/Hz higher than that before optimization on average, which verifies the effectiveness of the optimization. Secondly, we observe that when $N_b=16$, the curve ``Optimize RIS only" outperforms the curve ``Optimize Q only". However, when $N_b=32$ and $N_b=64$, the curve ``Optimize RIS only" performs worse than the curve ``Optimize Q only" and the gap enlarges. Thus, with the same number of antennas at the BS and reflection coefficients at the RIS, optimizing the transmit covariance matrix at the BS is superior to optimizing the reflection coefficients at the RIS. When the number of reflection coefficients is much larger than the number of antennas, optimizing the reflection coefficients outperforms optimizing the transmit covariance matrix. Thanks to the low cost of the RIS elements, the RIS is usually composed of a large number of reflection units. In this case, the systems still perform well when the CSI is not available at the BS and the reflection coefficients at the RIS is optimized.

Finally, Fig.~\ref{main_rislength} depicts the ergodic achievable rate against the number of antennas at the BS. The system parameter settings are the same as those of Fig.~\ref{cap_len}, while the proposed algorithm is adopted in Fig.~\ref{main_rislength}. Compared with Fig.~\ref{cap_len}, it is interesting to find that the ergodic achievable rate increases logarithmically with the number of antennas at the BS when the transmit covariance matrix is optimized. Thus, the optimization of the transmit covariance matrix is very important in the RIS-aided mmWave MIMO systems.

\section{Conclusion}
In this paper, we considered the ergodic achievable rate of an RIS-aided mmWave MIMO communication system under the SV channel model. The approximations of the ergodic achievable rate were derived by means of the majorization theory and Jensen's inequality. Based on the derived approximation, we maximized the ergodic achievable rate by jointly designing the transmit covariance and reflection coefficients at the BS and RIS, respectively. Numerical results validated the effectiveness of the proposed algorithms. It was found that the ergodic achievable rate after optimization can be improved by about 20 bps/Hz on average over a wide range of SNRs. In addition, when the transmitted signals on each antenna are i.i.d., the ergodic achievable rate remains unchanged after the number of antennas at the BS reaches a certain amount\footnote{This amount depends on the transmit SNR, the number of paths, the number of antennas at the user, and the number of reflection units at the RIS. When ${\mathbf{Q}}=\frac{P_T}{N_b} \mathbf{I}_{N_b}$,  $R_{Jen}(\mathbf{Q},\mathbf{\Theta})$ in Eq. \eqref{jen1} shows that increasing the number of antennas at the BS affects the ergodic achievable rate by influencing the eigenvalues of $\mathbf{A}_b^H \mathbf{A}_b$. However, when the number of antennas is large, the eigenvalues of $\mathbf{A}_b^H \mathbf{A}_b$ are nearly equal to one due to the asymptotic orthogonality of the array response vectors. Letting $\varpi_i=\frac{P_T N_u N_r^2}{\sigma^2 P L}  d_{u,i} d_{r,i}$, $R_{Jen}(\mathbf{Q},\mathbf{\Theta})$ will vary more rapidly with $N_b$ when $\{\varpi_i\}_{i=1,\ldots,N_s}$ are large. In the setting of Fig.~\ref{cap_len}, the amount of BS antennas beyond which  the ergodic achievable rate is nearly unchanged is about 80. }. Moreover, it was revealed that optimizing the reflection coefficients at the RIS is more effective than optimizing the transmit covariance matrix at the BS when there are a large number of reflection units. The analysis and results in this article provided insights on the deployment and achievable rate evaluation of RIS-assisted mmWave wireless systems. {Future work may consider the situation with multiple users and RISs.  When considering multiple users, there will be interference among different users.
The corresponding analysis and optimization of the achievable rate are likely to  be more difficult, thus new
challenges may arise and are worth in-depth investigation.}

\appendix
\begin{appendices}
\subsection{Proof of Lemma 1} \label{append1}
$R(\mathbf{Q}, \mathbf{\Theta})$ can be expressed as Eq. \eqref{ergo1},
\begin{figure*}
\begin{equation} \label{ergo1}
\begin{aligned}
R(\mathbf{Q}, \mathbf{\Theta}) &= \mathbb{E}_{\mathbf{H}}\left[\log _{2} \operatorname{det}\left(\mathbf{I}_{N_{u}}+\frac{1}{\sigma^{2}} \mathbf{H} \mathbf{Q H}^{H}\right)\right]\\
&= \mathbb{E}_{\mathbf{H}}\left[\log _{2} \operatorname{det}\left(\mathbf{I}_{N_{u}}
+ \frac{1}{\sigma^{2}} \mathbf{T} \mathbf{\Theta} \mathbf{G} \mathbf{Q} \mathbf{G}^H \mathbf{\Theta}^H \mathbf{T}^H \right)\right] \\
&= \mathbb{E}_{\mathbf{H}}\left[\log _{2} \operatorname{det}\left(\mathbf{I}_{N_{u}}
+ \frac{1}{\sigma^{2}} \mathbf{A}_u \mathbf{T}_l \mathbf{A}_{rl}^H \mathbf{\Theta} \mathbf{A}_{rp} \mathbf{G}_p \mathbf{A}_b^H
\mathbf{Q}  \mathbf{A}_b \mathbf{G}_p^H \mathbf{A}_{rp}^H \mathbf{\Theta}^H \mathbf{A}_{rl} \mathbf{T}_l^H \mathbf{A}_u^H
 \right)\right] \\
&= \mathbb{E}_{\mathbf{H}}\left[\log _{2} \operatorname{det}\left[ \prod \limits_{i=1}^{N_u} \left(1
+ \frac{1}{\sigma^{2}} \lambda_i\left( \mathbf{A}_u \mathbf{T}_l \mathbf{A}_{rl}^H \mathbf{\Theta} \mathbf{A}_{rp} \mathbf{G}_p \mathbf{A}_b^H
\mathbf{Q}  \mathbf{A}_b \mathbf{G}_p^H \mathbf{A}_{rp}^H \mathbf{\Theta}^H \mathbf{A}_{rl} \mathbf{T}_l^H \mathbf{A}_u^H \right) \right)
 \right]\right],
\end{aligned}
\end{equation}
\end{figure*}
where  $\lambda_i (\cdot)$ denotes the $i$-th largest eigenvalue of the input matrix.
According to the definition [\cite{marshall1979inequalities}, 1.A.7], $\forall \mathbf{x},\mathbf{y} \in \mathbb{R}^n_+$, $\mathbf{x}$ is said to be log-majorized by $\mathbf{y}$, denoted by $\mathbf{x} \prec_\text{log} \mathbf{y}$, if
\begin{equation}
\begin{split}
& \prod \limits_{i=1} ^k x_{[i]} \leq \prod \limits_{i=1}^k y_{[i]}, \quad k=1,\ldots,n-1, \\
& \prod \limits_{i=1} ^n x_{[i]} = \prod \limits_{i=1}^n y_{[i]}.
\end{split}
\end{equation}

According to [\cite{marshall1979inequalities}, Theorem 9.H.1.a] If $\mathbf{U}$ and $\mathbf{V}$ are $n\times n$ positive semidefinite Hermitian matrices, then
\begin{equation} \label{theorem9H}
\begin{split}
& \prod \limits_{i=1} ^k \lambda_i (\mathbf{UV}) \leq \prod \limits_{i=1}^k \lambda_i(\mathbf{U})\lambda_i(\mathbf{V}), \quad k=1,\ldots,n-1, \\
& \prod \limits_{i=1} ^n \lambda_i (\mathbf{UV}) = \prod \limits_{i=1}^n \lambda_i(\mathbf{U})\lambda_i(\mathbf{V}).
\end{split}
\end{equation}
Thus, $\boldsymbol{\lambda}  (\mathbf{UV}) \prec_\text{log} \boldsymbol{\lambda}(\mathbf{U}) \odot \boldsymbol{\lambda}(\mathbf{V})$, where $\odot$ denotes hadamard  product, and $\boldsymbol{\lambda}(\mathbf{Y}) = [\lambda_{1}(\mathbf{Y}), 
\lambda_2 (\mathbf{Y}), \ldots, \lambda_{n}(\mathbf{Y})]^T,  \mathbf{Y} \in \{\mathbf{UV}, \mathbf{U}, \mathbf{V}\}$. Therefore, we can obtain
\begin{small}
\begin{equation} \label{lambda_1}
\begin{aligned}
& \boldsymbol{\lambda}\left( \mathbf{A}_u \mathbf{T}_l \mathbf{A}_{rl}^H \mathbf{\Theta} \mathbf{A}_{rp} \mathbf{G}_p \mathbf{A}_b^H
\mathbf{Q}  \mathbf{A}_b \mathbf{G}_p^H \mathbf{A}_{rp}^H \mathbf{\Theta}^H \mathbf{A}_{rl} \mathbf{T}_l^H \mathbf{A}_u^H \right) \\
& \overset{(a)}{\prec}_\text{log} \boldsymbol{\lambda}(\mathbf{A}_u \mathbf{A}_u^H) \odot \boldsymbol{\lambda} \left( \mathbf{T}_l \mathbf{A}_{rl}^H \mathbf{\Theta} \mathbf{A}_{rp} \mathbf{G}_p \mathbf{A}_b^H
\mathbf{Q}  \mathbf{A}_b \mathbf{G}_p^H \mathbf{A}_{rp}^H \mathbf{\Theta}^H \mathbf{A}_{rl} \mathbf{T}_l^H  \right) \\
& \overset{(b)}{\prec}_\text{log} \boldsymbol{\lambda}(\mathbf{A}_u^H \mathbf{A}_u) \odot \boldsymbol{\lambda}(\mathbf{A}_b^H \mathbf{Q} \mathbf{A}_b) \odot
 \boldsymbol{\lambda}(\mathbf{X}^H \mathbf{X}) \odot \boldsymbol{\lambda}(\mathbf{G}_p^H \mathbf{G}_p) \odot
 \boldsymbol{\lambda}(\mathbf{T}_l^H \mathbf{T}_l),
\end{aligned}
\end{equation}
\end{small}where $(a)$ follows $\boldsymbol{\lambda}(\mathbf{AB}) = \boldsymbol{\lambda} (\mathbf{BA})$,  $(b)$ follows applying Theorem in \eqref{theorem9H} repeatedly, and $\mathbf{X} = \mathbf{A}_{rl}^H \mathbf{\Theta} \mathbf{A}_{rp}$.

According to definition [\cite{marshall1979inequalities}, 1.A.2], $\forall \mathbf{x},\mathbf{y} \in \mathbb{R}^n$, $\mathbf{x}$ is said to be weakly majorized by $\mathbf{y}$, denoted by $\mathbf{x} \prec_w \mathbf{y}$, if
\begin{equation}
\sum \limits_{i=1}^k x_{[i]} \leq \sum \limits_{i=1}^k y_{[i]},\quad k=1, \ldots, n-1.
\end{equation}

According to [\cite{zhan2004matrix}, Theorem 2.7], Let the components of $\mathbf{x},\mathbf{y} \in \mathbb{R}^n$ be nonnegative. Then
\begin{equation}
\mathbf{x} \prec_{\text{log}} \mathbf{y} \quad \text{implies} \quad \mathbf{x} \prec_w \mathbf{y}.
\end{equation}

Thus, \eqref{lambda_1} can be further simplified as
\begin{small}
\begin{equation}
\begin{aligned}
& \boldsymbol{\lambda}\left( \mathbf{A}_u \mathbf{T}_l \mathbf{A}_{rl}^H \mathbf{\Theta} \mathbf{A}_{rp} \mathbf{G}_p \mathbf{A}_b^H
\mathbf{Q}  \mathbf{A}_b \mathbf{G}_p^H \mathbf{A}_{rp}^H \mathbf{\Theta}^H \mathbf{A}_{rl} \mathbf{T}_l^H \mathbf{A}_u^H \right) \\
& {\prec}_w \boldsymbol{\lambda}(\mathbf{A}_u^H \mathbf{A}_u) \odot \boldsymbol{\lambda}(\mathbf{A}_b^H \mathbf{Q} \mathbf{A}_b) \odot
 \boldsymbol{\lambda}(\mathbf{X}^H \mathbf{X}) \odot \boldsymbol{\lambda}(\mathbf{G}_p^H \mathbf{G}_p) \odot
 \boldsymbol{\lambda}(\mathbf{T}_l^H \mathbf{T}_l).
\end{aligned}
\end{equation}
\end{small}

According to [\cite{marshall1979inequalities}, 1.A.1], $\forall \mathbf{x},\mathbf{y} \in \mathbb{R}^n$, $\mathbf{x}$ is said to be majorized by $\mathbf{y}$, denoted by $\mathbf{x} \prec  \mathbf{y}$, if
\begin{equation}
\begin{split}
& \sum \limits_{i=1} ^k x_{[i]} \leq \sum \limits_{i=1}^k y_{[i]}, \quad k=1,\ldots,n-1, \\
& \sum \limits_{i=1} ^n x_{[i]} = \sum \limits_{i=1}^n y_{[i]}.
\end{split}
\end{equation}

According to [\cite{marshall1979inequalities}, 3.C.1], if $\mathbf{I} \subset \mathbb{R}$ is an interval and $g: \mathbf{I} \rightarrow \mathbb{R}$ is convex, then
\begin{equation} \label{theorem_concave}
\phi(\mathbf{x}) = \sum \limits_{i=1}^n g(x_i)
\end{equation}
is Schur-convex on $\mathbf{I}^n$. Consequently, $\mathbf{x} \prec \mathbf{y}$ on $\mathbf{I}^n$ implies $\phi(\mathbf{x})\leq \mathbf{y}$. Similarly, if $g$ is concave, the $\phi(\mathbf{x})$ is Schur-concave.

According to [\cite{marshall1979inequalities}, 5.A.9], $\forall \mathbf{x},\mathbf{y} \in \mathbb{R}^n$, if $\mathbf{x} \prec_w \mathbf{y}$, then there exists a vector $\mathbf{u}$ such that $\mathbf{x}\leq \mathbf{u}$ and $\mathbf{u} \prec \mathbf{y}$.

Thus, there exists a vector $\mathbf{u}$ such that Eq. \eqref{append_u} can be satisfied.
\begin{figure*}
\begin{subequations} \label{append_u}
\begin{align}
& \boldsymbol{\lambda}\left( \mathbf{A}_u \mathbf{T}_l \mathbf{A}_{rl}^H \mathbf{\Theta} \mathbf{A}_{rp} \mathbf{G}_p \mathbf{A}_b^H
\mathbf{Q}  \mathbf{A}_b \mathbf{G}_p^H \mathbf{A}_{rp}^H \mathbf{\Theta}^H \mathbf{A}_{rl} \mathbf{T}_l^H \mathbf{A}_u^H \right) \leq \mathbf{u}, \label{u1} \\
& \mathbf{u} \prec \boldsymbol{\lambda}(\mathbf{A}_u^H \mathbf{A}_u) \odot \boldsymbol{\lambda}(\mathbf{A}_b^H \mathbf{Q} \mathbf{A}_b) \odot
 \boldsymbol{\lambda}(\mathbf{X}^H \mathbf{X}) \odot \boldsymbol{\lambda}(\mathbf{G}_p^H \mathbf{G}_p) \odot
 \boldsymbol{\lambda}(\mathbf{T}_l^H \mathbf{T}_l)
\end{align}
\end{subequations}
\end{figure*}
Define $g(\lambda)= \log _{2} (1+\frac{1}{\sigma^2} \lambda)$ and $\phi(\boldsymbol{\lambda}) = \sum \limits_{i=1}^{N_s} g(\lambda_i)$, then $g(\lambda)$ is an increasing concave function. According to \eqref{theorem_concave}, $\phi(\boldsymbol{\lambda})$ is an increasing Schur-concave function. Then, we have Eq. \eqref{u2} and \eqref{u3}.
\begin{figure*}
\begin{equation} \label{u2}
\phi\left(\boldsymbol{\lambda}\left( \mathbf{A}_u \mathbf{T}_l \mathbf{A}_{rl}^H \mathbf{\Theta} \mathbf{A}_{rp} \mathbf{G}_p \mathbf{A}_b^H
\mathbf{Q}  \mathbf{A}_b \mathbf{G}_p^H \mathbf{A}_{rp}^H \mathbf{\Theta}^H \mathbf{A}_{rl} \mathbf{T}_l^H \mathbf{A}_u^H \right)\right) \leq \phi(\mathbf{u}).
\end{equation}
\begin{equation} \label{u3}
\phi( \boldsymbol{\lambda}(\mathbf{A}_u^H \mathbf{A}_u) \odot \boldsymbol{\lambda}(\mathbf{A}_b^H \mathbf{Q} \mathbf{A}_b) \odot
 \boldsymbol{\lambda}(\mathbf{X}^H \mathbf{X}) \odot \boldsymbol{\lambda}(\mathbf{G}_p^H \mathbf{G}_p) \odot
 \boldsymbol{\lambda}(\mathbf{T}_l^H \mathbf{T}_l) ) \leq \phi(\mathbf{u}).
\end{equation}
\end{figure*}
Therefore, combining \eqref{u2} and \eqref{u3}, we have Eq. \eqref{append_phi}.
\begin{figure*}
\begin{equation} \label{append_phi}
\begin{split}
 & \phi\left(\boldsymbol{\lambda}\left( \mathbf{A}_u \mathbf{T}_l \mathbf{A}_{rl}^H \mathbf{\Theta} \mathbf{A}_{rp} \mathbf{G}_p \mathbf{A}_b^H
\mathbf{Q}  \mathbf{A}_b \mathbf{G}_p^H \mathbf{A}_{rp}^H \mathbf{\Theta}^H \mathbf{A}_{rl} \mathbf{T}_l^H \mathbf{A}_u^H \right)\right) \\
&\approx
\phi\left(\boldsymbol{\lambda}(\mathbf{A}_u^H \mathbf{A}_u) \odot \boldsymbol{\lambda}(\mathbf{A}_b^H \mathbf{Q} \mathbf{A}_b) \odot
 \boldsymbol{\lambda}(\mathbf{X}^H \mathbf{X}) \odot \boldsymbol{\lambda}(\mathbf{G}_p^H \mathbf{G}_p) \odot
 \boldsymbol{\lambda}(\mathbf{T}_l^H \mathbf{T}_l)\right) .
 \end{split}
\end{equation}
\end{figure*}
Assume the eigenvalue decompositions (EVD) of $\mathbf{A}_u^H \mathbf{A}_u$, $\mathbf{A}_b^H \mathbf{Q} \mathbf{A}_b$ and $\mathbf{X}^H \mathbf{X}$ can be expressed as following,
\begin{equation}
\begin{split}
 & \mathbf{A}_u^H \mathbf{A}_u = \mathbf{U}_{u}^H \mathbf{D}_u \mathbf{U}_{u}, \\
& \mathbf{A}_b^H \mathbf{Q}\mathbf{A}_b = \mathbf{U}_{b}^H \mathbf{D}_b \mathbf{U}_{b}, \\
& \mathbf{X}^H \mathbf{X} = \mathbf{U}_{r}^H \mathbf{D}_r \mathbf{U}_{r},
 \end{split}
\end{equation}
where $\mathbf{U}_{u}$, $\mathbf{U}_{b}$, and $\mathbf{U}_{r}$ are the eigenvectors of $\mathbf{A}_u^H \mathbf{A}_u$, $\mathbf{A}_b^H \mathbf{Q}\mathbf{A}_b$, and $\mathbf{X}^H \mathbf{X}$, respectively; $ \mathbf{D}_u = \operatorname{diag} (d_{u,1}, d_{u,2}, \ldots, d_{u,L})$, $\mathbf{D}_b = \operatorname{diag} (d_{b,1}, d_{b,2}, \ldots, d_{b,P})$, $\mathbf{D}_r = \operatorname{diag} (d_{r,1}, d_{r,2}, \ldots, d_{r,P})$, and $d_{u,i}, d_{b,i}, d_{r,i}\geq 0$  are the eigenvalue of $\mathbf{A}_u^H \mathbf{A}_u$, $\mathbf{A}_b^H \mathbf{Q}\mathbf{A}_b$, and $\mathbf{X}^H \mathbf{X}$ in descending order, respectively.

Then, the ergodic achievable rate $R(\mathbf{Q}, \mathbf{\Theta})$ in Eq. \eqref{ergo1} can be approximated by Eq. \eqref{append_r_bo},
\begin{figure*}
\begin{equation} \label{append_r_bo}
\begin{aligned}
\widetilde{R}(\mathbf{Q}, \mathbf{\Theta})
& = \mathbb{E}_{\lambda} \left[\log _{2} \operatorname{det}\left(\mathbf{I}_{N_{s}} + \frac{1}{\sigma^2} \boldsymbol{\lambda}(\mathbf{A}_u^H  \mathbf{A}_u) \odot \boldsymbol{\lambda}(\mathbf{A}_b^H \mathbf{Q}\mathbf{A}_b) \odot
 \boldsymbol{\lambda}(\mathbf{X}^H \mathbf{X}) \odot \boldsymbol{\lambda}(\mathbf{G}_p^H \mathbf{G}_p) \odot
 \boldsymbol{\lambda}(\mathbf{T}_l^H \mathbf{T}_l) \right)\right] \\
& \overset{(a)}{=} \mathbb{E}_{g,t} \left[\sum \limits_{i=1}^{N_s} \log _{2} \left( 1+ \frac{ N_b N_u N_r^2 }{\sigma^2 P L } d_{b,i} d_{u,i} d_{r,i} |g_i|^2 |t_i|^2  \right) \right],
\end{aligned}
\end{equation}
\end{figure*}
where $N_s= \min\left(\operatorname{rank}(\mathbf{A}_b^H \mathbf{Q} \mathbf{A}_b), \operatorname{rank}(\mathbf{A}_u^H \mathbf{A}_u), \operatorname{rank}(\mathbf{X}^H \mathbf{X})\right)$ denotes the data streams, $(a)$ follows $ \boldsymbol{\lambda}(\mathbf{G}_p^H \mathbf{G}_p) = \frac{N_b N_r}{P} \left[|g_1|^2, |g_2|^2, \ldots, |g_P|^2\right]^T$ and $ \boldsymbol{\lambda}(\mathbf{T}_l^H \mathbf{T}_l) = \frac{N_u N_r}{L} \left[|t_1|^2, |t_2|^2, \ldots, |T_L|^2\right]^T$.

\subsection{Proof of Corollary 1} \label{proof_pro1}
Since $\mathbf{A}_b$, $\mathbf{A}_u$, $\mathbf{A}_{rp}$ and $\mathbf{A}_{rl}$ are composed of the columns of unitary matrices, the AoA at the RIS is symmetric to the AoD at the RIS, $\mathbf{Q} = \frac{P_T}{N_b} \mathbf{I}_{N_b}$, and $\mathbf{\Theta}=\mathbf{I}_{N_r}$, then $\mathbf{A}_b^H \mathbf{Q} \mathbf{A}_b = \frac{P_T}{N_b} \mathbf{I}_P$, $\mathbf{A}_u^H \mathbf{A}_u = \mathbf{I}_L$ and $\mathbf{X} = \mathbf{A}_{rl}^H \mathbf{\Theta} \mathbf{A}_{rp} = \mathbf{I}_{L \times P}$. Thus, the ergodic achievable rate in \eqref{capacity_ori} can be expressed as Eq. \eqref{append_rate},
\begin{figure*}
\begin{equation} \label{append_rate}
\begin{aligned}
{R}(\mathbf{Q}, \mathbf{\Theta}) &= \mathbb{E}_{\mathbf{H}}\left[\log _{2} \operatorname{det}\left(\mathbf{I}_{N_{u}}+\frac{1}{\sigma^{2}} \mathbf{H} \mathbf{Q H}^{H}\right)\right]\\
&= \mathbb{E}_{\mathbf{G}_p, \mathbf{T}_L}\left[\log _{2} \operatorname{det}\left(\mathbf{I}_{N_{u}}
+ \frac{1}{\sigma^{2}} \mathbf{A}_u \mathbf{T}_l \mathbf{A}_{rl}^H \mathbf{\Theta} \mathbf{A}_{rp} \mathbf{G}_p \mathbf{A}_b^H
\mathbf{Q}  \mathbf{A}_b \mathbf{G}_p^H \mathbf{A}_{rp}^H \mathbf{\Theta}^H \mathbf{A}_{rl} \mathbf{T}_l^H \mathbf{A}_u^H
 \right)\right] \\
&= \mathbb{E}_{\mathbf{G}_p, \mathbf{T}_L}\left[\log _{2} \operatorname{det}\left(\mathbf{I}_{N_{u}}
+ \frac{P_T}{\sigma^{2} N_b} \mathbf{I}_{L\times P} \mathbf{G}_p \mathbf{G}_p^H \mathbf{I}_{P\times L} \mathbf{T}_l^H \mathbf{T}_l
 \right)\right] \\
& \overset{(a)}{=} \mathbb{E}_{g,t} \left[\sum \limits_{i=1}^{N_s} \log _{2} \left( 1+ \frac{ P_T N_u N_r^2 }{\sigma^2 P L } |g_i|^2 |t_i|^2  \right) \right],
\end{aligned}
\end{equation}
\end{figure*}
where $(a)$ holds due to $\mathbf{G}_p = \sqrt{\frac{N_r N_{b}}{P}} \text{diag}( g_1, g_2, \ldots, g_P)$ and $ \mathbf{T}_l = \sqrt{\frac{N_r N_{u}}{L}} \text{diag}( t_1, t_2, \ldots, t_L)$. At the same time, we have $d_{b,i} =\frac{P_T}{N_b}, d_{u,i}=1, d_{r,i} = 1, i=1,2,\ldots, N_s$ due to the fact that $\mathbf{A}_b^H \mathbf{Q} \mathbf{A}_b = \frac{P_T}{N_b} \mathbf{I}_P$, $\mathbf{A}_u^H \mathbf{A}_u = \mathbf{I}_L$ and $\mathbf{X} = \mathbf{A}_{rl}^H \mathbf{\Theta} \mathbf{A}_{rp} = \mathbf{I}_{L \times P}$. Therefore, the approximation in Eq. \eqref{app_ori} can be rewritten as
\begin{equation}
\widetilde{R}(\mathbf{Q}, \mathbf{\Theta})
= \mathbb{E}_{g,t} \left[\sum \limits_{i=1}^{N_s} \log _{2} \left( 1+ \frac{ P_T N_u N_r^2 }{\sigma^2 P L } |g_i|^2 |t_i|^2  \right) \right].
\end{equation}
Consequently, the derived approximation is identical to the exact ergodic achievable rate.

\subsection{Proof of Corollary 2} \label{appendix_theorem2}
Since $g_i \sim \mathcal{C}\mathcal{N} (0,1)$, $t_i \sim \mathcal{C}\mathcal{N} (0,1)$, then $|g_i|^2 \sim \exp (1)$ and $|t_i|^2 \sim \exp (1)$. Let $ x = |g_i|^2 |t_i|^2$, the probability density function (PDF) $f_X(x)$ can be expressed as
\begin{equation} \label{gailv_fx}
f_X(x) = \int_0^\infty \frac{1}{z} e^{-(z+x/z)} dz, \quad x>0.
\end{equation}

Then, the approximation $\widetilde{R}(\mathbf{Q}, \mathbf{\Theta})$ can be calculated as
\begin{small}
\begin{equation}
\begin{aligned}
\widetilde{R}(\mathbf{Q}, \mathbf{\Theta})
& =  \mathbb{E}_{g,t} \left[\sum \limits_{i=1}^{N_s} \log _{2} \left( 1+ \frac{ N_b N_u N_r^2 }{\sigma^2 P L } d_{b,i} d_{u,i} d_{r,i} |g_i|^2 |t_i|^2  \right) \right] \\
& \overset{(a)}{=} \sum \limits_{i=1}^{N_s} \int _{0}^{\infty} f_X (x) \log_2 \left( 1+ \alpha_i x \right) dx \\
& \overset{(b)}{=}\sum \limits_{i=1}^{N_s} \int _{0}^{\infty} \int _{0}^{\infty} \frac{1}{z} e^{-z+x/z} \log_2 \left( 1+ \alpha_i x \right) dx dz \\
& = \frac{1}{ \ln 2} \sum \limits_{i=1}^{N_s} \int_{0}^\infty
e^{-z} \left( \int _{0}^{\infty} \frac{1}{z} e^{-x/z} \ln \left(1+\alpha_i z\right) dx \right)  dz \\
& \overset{(c)}{=} \frac{1}{ \ln 2} \sum \limits_{i=1}^{N_s} \int_{0}^\infty
e^{-z} e^{ \frac{1}{\alpha_i z} } E_1 \left(\frac{1}{\alpha_i z} \right) dz,
\end{aligned}
\end{equation}
\end{small}where $(a)$ holds by letting $\alpha_i= \frac{ N_b N_u N_r^2 }{\sigma^2 P L } d_{b,i} d_{u,i} d_{r,i}$,  $(b)$ follows Eq. \eqref{gailv_fx}, and $(c)$ holds due to the fact that $\int_{0}^\infty e^{-x} \ln \left(1+ \frac{x}{\alpha}\right) dx= e^\alpha E_1(\alpha)$ \cite{8816689}.

\subsection{Proof of Lemma 2} \label{append2}
If $c_i=0$, we can add a very small real number $\delta>0$, i.e., $c_i = \delta$. Then, we have Eq. \eqref{prove_1},
\begin{figure*}
\begin{equation} \label{prove_1}
\begin{split}
  \operatorname{det} \left[ \mathbf{I} + \operatorname{diag}(\mathbf{c}) \mathbf{U}^H \operatorname{diag}(\mathbf{s}) \mathbf{U}\right]
&= \operatorname{det} \left[\operatorname{diag} (\mathbf{c})\left( \operatorname{diag} (  \mathbf{c}^{-1})  + \mathbf{U}^H \operatorname{diag} (\mathbf{s})\mathbf{U} \right)\right]\\
 &= \operatorname{det} \left[\operatorname{diag} (\mathbf{c})\left(\mathbf{U} \operatorname{diag} (  \mathbf{c}^{-1}) \mathbf{U}^H + \operatorname{diag} (\mathbf{s}) \right)\right]\\
 &\overset{(a)}{=} \operatorname{det} \left[\operatorname{diag} (\mathbf{c})\right] \operatorname{det} \left[ \mathbf{U} \operatorname{diag} (  \mathbf{c}^{-1}) \mathbf{U}^H + \operatorname{diag} (\mathbf{s}) \right],
 \end{split}
\end{equation}
\end{figure*}
where $(a)$ holds due  to $\operatorname{det}(\mathbf{AB})= \operatorname{det} (\mathbf{A}) \operatorname{det} (\mathbf{B})$ for arbitrary matrices $\mathbf{A}$ and $\mathbf{B}$.

According to [\cite{marshall1979inequalities}, 9.G.3], if $\mathbf{G}$ and $\mathbf{H}$ are $n\times n$ Hermitian matrices, then
\begin{equation}
 \operatorname{det}(\mathbf{G}+\mathbf{H})=\prod_{i=1}^{n} \lambda_{i}(\mathbf{G}+\mathbf{H}) \leq \prod_{i=1}^{n}\left[\lambda_{i}(\mathbf{G})+\lambda_{n-i+1}(\mathbf{H})\right],
\end{equation}
where $\lambda_i (\cdot)$ denotes the $i$-th largest eigenvalue of the input matrix. Thus, we have Eq. \eqref{prove_3},
\begin{figure*}
\begin{equation} \label{prove_3}
\begin{split}
\operatorname{det} \left[ \mathbf{U} \operatorname{diag} (  \mathbf{c}^{-1}) \mathbf{U}^H + \operatorname{diag} (\mathbf{s}) \right]
& = \prod_{i=1}^{N} \lambda_{i} \left( \mathbf{U} \operatorname{diag} (  \mathbf{c}^{-1}) \mathbf{U}^H + \operatorname{diag} (\mathbf{s}) \right) \\
& \overset{(a)}{\leq} \prod_{i=1}^{N}\left[\lambda_{i}(\mathbf{U} \operatorname{diag} (  \mathbf{c}^{-1}) \mathbf{U}^H) + \lambda_{N-i+1}(\operatorname{diag} (\mathbf{s})) \right] \\
& \overset{(b)}{=} \prod_{i=1}^{N}\left[\lambda_{i}( \operatorname{diag} (  \mathbf{c}^{-1}) ) + \lambda_{N-i+1}(\operatorname{diag} (\mathbf{s})) \right] \\
& \overset{(c)}{=} \prod_{i=1}^N \left( c_{N-i+1}^{-1} + s_{N-i+1} \right) ,
\end{split}
\end{equation}
\end{figure*}
where $(a)$ holds due to the Theorem [\cite{marshall1979inequalities}, 9.G.3], $(b)$ holds due to $\lambda (\mathbf{AB}) = \lambda(\mathbf{BA})$ for arbitrary matrices $\mathbf{A}$ and $\mathbf{B}$, and $(c)$ holds due to the fact that $\mathbf{c}$ and $\mathbf{s}$ are the descending ordered sequences.
As a result, we have Eq. \eqref{append_det}.
\begin{figure*}
\begin{equation} \label{append_det}
\begin{split}
  \operatorname{det} \left[ \mathbf{I} + \operatorname{diag}(\mathbf{c}) \mathbf{U}^H \operatorname{diag}(\mathbf{s}) \mathbf{U}\right]
 &= \operatorname{det} \left[\operatorname{diag} (\mathbf{c})\right] \operatorname{det} \left[ \mathbf{U} \operatorname{diag} (  \mathbf{c}^{-1}) \mathbf{U}^H + \operatorname{diag} (\mathbf{s}) \right] \\
 & \leq \prod \limits_{i=1}^N c_i \prod \limits_{i=1}^N (c_i^{-1} + s_i) \\
 & = \prod \limits_{i=1}^N (1+c_i s_i) \\
 & = \operatorname{det}   \left[ \mathbf{I} + \mathbf{U}^H  \operatorname{diag}(\mathbf{c}) \operatorname{diag}(\mathbf{s}) \mathbf{U}\right].
 \end{split}
\end{equation}
\end{figure*}

\end{appendices}

\bibliographystyle{IEEEtran}
\bibliography{Erg_Cap}

\end{document}